\documentclass[11pt]{article}

\usepackage{graphics,latexsym}
\usepackage{graphicx}
\usepackage{amsmath,bm,graphicx,epsfig}
\usepackage{amssymb}

\begin{document}

\centerline{\Large \bf ROTATING ELECTROMAGNETIC WAVES}
\vspace{.2cm}
\centerline{\Large \bf IN TOROID-SHAPED REGIONS}
\vspace{.7cm}
\centerline{CLAUDIA CHINOSI}

\centerline{Dipartimento di Scienze e Tecnologie Avanzate}

\centerline{Universit\`a del Piemonte Orientale}
 
\centerline{Viale Teresa Michel 11, 15121 Alessandria, Italy}

\centerline{chinosi@unipmn.it}

\vspace{.5cm}

\centerline{LUCIA DELLA CROCE}

\centerline{Dipartimento di Matematica}

\centerline{Universit\`a di Pavia}

\centerline{Via Ferrata 1, 27100 Pavia, Italy}

\centerline{lucia.dellacroce@unipv.it}

\vspace{.5cm}

\centerline{DANIELE FUNARO}

\centerline{Dipartimento di Matematica}

\centerline{Universit\`a di Modena e Reggio Emilia}

\centerline{Via Campi 213/B, 41125 Modena, Italy}
 
\centerline{daniele.funaro@unimore.it}

\begin{abstract}
Electromagnetic waves, solving the full set  of Maxwell  equations
in vacuum, are numerically computed. These waves occupy a fixed
bounded region of the three dimensional space, topologically
equivalent to a toroid. Thus, their fluid dynamics analogs are
vortex rings. An analysis of the shape of the sections of the
rings, depending on the angular speed of rotation and the major
diameter, is carried out. Successively, spherical electromagnetic
vortex rings of Hill's type are taken into consideration. For some
interesting peculiar configurations, explicit numerical solutions
are exhibited.
\end{abstract}

\vspace{.8cm}
\noindent Keywords: Electromagnetism; Solitary wave; Optimal set; Toroid; Hill's vortex.

\vspace{.2cm}
\noindent {PACS Nos.: 41.20.Jb; 47.32.Ef; 02.70.Dh}

\vspace{.2cm}
\noindent{Published in the International Journal of Modern Physics C, Vol. 21, 
No. 1 (2010), pp. 11-32. DOI: 10.1142/S0129183110014926}

\section{Introduction}\label{sec1}

 We start by recalling that the classical set of Maxwell equations in vacuum has the
form:
\begin{equation}\label{eq:rotb}
 \frac{\partial {\bf E}}{ \partial t}~=~ c^2 {\rm curl} {\bf B}
\end{equation}
\begin{equation}\label{eq:dive}
{\rm div}{\bf E} ~=~0
\end{equation}
\begin{equation}\label{eq:rote}
\frac{\partial {\bf B}}{\partial t}~=~ -{\rm curl} {\bf E}
\end{equation}
\begin{equation}\label{eq:divb}
{\rm div}{\bf B} ~=~0
\end{equation}
where $c$ is the speed of light and the two fields ${\bf E}$ and $c{\bf B}$ have the same dimensions.
\par\smallskip

It is customary to introduce the potentials ${\bf A}$ and $\Phi$ such that:
\begin{equation}\label{eq:potenz}
{\bf B}~=~\frac {1}{c}~{\rm curl}{\bf A}~~~~~~~~~{\bf E}~=~-\frac {1}{c} \frac {\partial {\bf A}}
{\partial t}~-~\nabla \Phi
\end{equation}
By this assumption, equations (\ref{eq:rote}) and (\ref{eq:divb}) are automatically satisfied. In
addition, we require that the vector and the scalar potentials are related by the following Lorenz
gauge condition:
\begin{equation}\label{eq:color}
{\rm div}{\bf A}~+~\frac {1}{c} \frac {\partial \Phi}{\partial t}~=~0
\end{equation}
Furthermore, one can deduce the two wave equations:
\begin{equation}\label{eq:ondea}
\frac {\partial^2 {\bf A}} {\partial t^2}~-~c^2\Delta {\bf A}~=~0
\end{equation}
\begin{equation}\label{eq:ondea1}
\frac {\partial^2\Phi} {\partial t^2}~-~c^2\Delta \Phi~=~0
\end{equation}
\par\smallskip

We are concerned with studying, from the numerical  viewpoint, the
development of a solitary electromagnetic wave, trapped in a
bounded region of space having a toroid shape. In the cylindrical
case (equivalent to a 2-D problem), and partly for 3-D problems,
this analysis was proposed and carried out in  Ref.~\cite{funaro}.
There, the goal was to simulate stable elementary subatomic
particles by means of rotating photons. Here, we would like to
continue the discussion of the 3-D case, because the subject might be
of more general interest. Since exact solutions are only available
in special situations, we shall make use of finite
element techniques to approximate the model equations. In order to
determine solutions of the vector wave equation, confined in
appropriate steady domains, we will perform an in-depth analysis
of the lower spectrum of a suitable elliptic operator, in
dependance of the shape and magnitude of the regions. In
particular, we will be concerned with those domains realizing the
coincidence of the fourth and the fifth eigenvalues of the
differential operator. The reasons for this choice will become
clear to the reader as we proceed with the investigation.

\par\smallskip

The structures we consider in this paper display a strong analogy
with fluid dynamics vortex rings (see for instance Ref.~\cite{batchelor},
\cite{lim}, \cite{shariff}). For this reason,
the techniques we apply here may be useful to get additional
results in the study of the development of fluid vortices. As a
matter of fact, peculiar configurations are going to be presented
and discussed, opening the path to an interesting scenario for
future extensions.

\section{The Cylindrical Case}

Let us quickly review the results obtained in Ref.~\cite{funaro},
concerning waves rotating around the axis of a cylinder. The
magnetic field is oriented in the same direction  as the $z$-axis,
so that the electric field lays on the orthogonal plane. The
variables are expressed in cylindrical coordinates $(r, z, \phi)$.
The solutions, however, will not depend on $z$.
\par\smallskip

Several options  are examined in Ref.~\cite{funaro}, here we show the
most significant one, that will be used later to construct the
toroid case. We denote by $\omega$ a positive parameter which
characterizes the frequency of rotation of the wave:  $\nu
=c\omega /\pi$. Up to multiplicative constant, the two vector
fields are given by:
$${\bf E}=\left(
\frac {2 J_2(\omega r)} {\omega r}\cos (c\omega t-2\phi) ,~~0,~~ J_2^\prime (\omega r)\sin (c\omega
t-2\phi)\right)$$
\begin{equation}\label{eq:cbdisk}
{\bf B}~=~\frac {1} {c}\Big( 0,~~  J_2(\omega r) \cos (c\omega t-2\phi),~~0\Big)
\end{equation}
for $0\leq \phi < 2\pi$, $0\leq r\leq \delta_0 /\omega ~$ and any
$z$.  The fields in (\ref{eq:cbdisk}) are defined on a disk
of radius $\delta_0 /\omega$, whose size is inversely proportional
to the angular speed of rotation. In Ref.~\cite{funaro} , figure 5.7,
the reader can see the displacement of the electric field for
$t=0$ (some on-line animations can be viewed in Ref.~\cite{funweb}, {\it rotating photons}).
\par\smallskip

In (\ref{eq:cbdisk}) we find the Bessel function $J_k$ (see, e.g., Ref.~\cite{watson}) that solves the
following differential equation:
\begin{equation}\label{eq:bes4}
J_k^{\prime\prime}(x)~+~\frac {J_k^\prime (x)} { x}~-~\frac{k^2J_k(x)} {x^2}~+~ J_k(x)~=~0
\end{equation}
It is also useful to recall that Bessel functions are connected by the relations:
\begin{equation}\label{eq:besre}
J_k^\prime (x) ~+~\frac {k J_k(x)} {x}~=~ J_{k-1}(x)
\end{equation}
\begin{equation}\label{eq:besre2}
J_{k+1}(x) ~=~\frac {2k J_k(x)} {x}~-~ J_{k-1}(x)
\end{equation}
 The quantity $~\delta_0\approx
5.135622~$ turns out to be the first zero of $J_2$.  In this way,
for $r=\delta_0 /\omega $, the components  $E_1$ and $B_2$ are
zero. Note also that ${\bf E}$ and ${\bf B}$ vanish for $r=0$,
since $J_k(x)$ decays as $x^k$ for $x\rightarrow 0$. The choice
$k=1$ is not permitted because it does not allow to prolong with
continuity the fields up to $r=0$, although, one could take into
consideration solutions for $k$ integer greater than 2:
$${\bf E}=\left(
\frac {k J_k(\omega r)} {\omega r}\cos (c\omega t-k\phi) ,~~0,~~ J_k^\prime (\omega r)\sin (c\omega
t-k\phi)\right)$$
\begin{equation}\label{eq:cbdiskk}
{\bf B}~=~\frac {1} {c}\Big( 0,~~  J_k(\omega r) \cos (c\omega t-k\phi),~~0\Big)
\end{equation}
The idea is to simulate a $k$-body rotating system in equilibrium.
This somehow explains why the case $k=1$ is not going to produce
meaningful solutions.
\par\smallskip

For $k=2$, the electromagnetic fields in (\ref{eq:cbdisk})  are
generated  by the following potentials:
$${\bf A}~=~ -\frac {1} {\omega} \Big(J_3(\omega r)\sin (c\omega t-2\phi), ~~0~,
~~  J_3(\omega r)\cos (c\omega t-2\phi)\Big)$$
$$ \Phi ~=~-\frac {1} {\omega}
 J_2(\omega r)\cos (c\omega t-2\phi)$$
satisfying the Lorenz condition (\ref{eq:color}) and the equations
(\ref{eq:ondea})-(\ref{eq:ondea1}). The reader can check this by
direct differentiation. We observe that $A_1$ and $A_3$ have a
phase difference of 45 degrees, since: $~\sin (c\omega t-2\phi )=
\cos (c\omega t-2(\phi +\pi /4))$.
Denoting by $a(r)=J_3(\omega r)$
the term depending only on the radial variable, we easily discover
that (see (\ref{eq:bes4}) for $k=3$):
\begin{equation}\label{eq:eqlap}
-\left(\frac {d^2} {dr^2}+\frac {1} {r}\frac {d}  {dr}-\frac {9} {r^2}\right) a~=~\omega^2 a
\end{equation}
We can introduce a new  function $w=\frac {d} {dr}a+\frac {3}
{r}a$. A straightforward computation, using relation (\ref{eq:eqlap})
and its derivative, brings to:
$$
-\left(\frac {d^2} {dr^2}+\frac {1} {r}\frac {d}  {dr}-\frac {4} {r^2}\right) w~=~
-\left( a^{\prime\prime\prime}+\frac{a^{\prime\prime}}{r}-\frac{10a^\prime}{r^2}
+\frac{18a}{r^3}\right)
$$
\begin{equation}\label{eq:eqlapp}
-~\frac{3}{r}\left(a^{\prime\prime}+\frac {a^\prime}{r}-\frac{9a}{r^2}
\right)~=~\omega^2\left( a^\prime +\frac{3a}{r}\right)~=~\omega^2 w
\end{equation}

By scaling the interval in order to impose the boundary
conditions $w(0)=0$ and $w(1)=0$, the eigenvalue problem
(\ref{eq:eqlapp}) takes the form:
\begin{equation}\label{eq:eqlap2}
Lw~=~-\left(\frac {d^2} {dr^2}+\frac {1} {r}\frac {d}  {dr}-\frac {4} {r^2}\right) w~=~\delta_0^2 w
\end{equation}
where $L$ is a positive-definite differential operator.  Such  an
operator is the same as the one we would obtain from the
Laplacian, after separation of variables in polar coordinates, by
imposing homogeneous Dirichlet boundary conditions on a disk
$\Omega$ (on the plane $(r,\phi )$) of radius equal to 1. In this
circumstance, the eigenvalue:
\begin{equation}\label{eq:auto2d}
\lambda ~=~\delta_0^2~\approx~26.37461
\end{equation}
has double multiplicity. The corresponding eigenfunctions are
orthogonal in $L^2(\Omega )$ and show a phase difference of 45
degrees (see figures 1 and 2).

\vspace{.6cm}

\begin{table}[h!]
\label{tab1} \noindent\[
\begin{array}{|c|c|}
  \hline
  ~~~~~~~\lambda~~~~~~~ &   {\it multiplicity}  \\
  \hline
  ~~5.78318 & 1 \\
  14.68197 & 2 \\
  26.37461 & 2 \\
  30.47126 & 1 \\
  40.70646 & 2 \\
  49.21845 & 2 \\
  57.58294 & 2 \\
  70.84999 & 2 \\
    \hline
\end{array}
\]
\caption{\small Eigenvalues counted with their
multiplicity of the Laplacian on a disk of radius 1, with
homogeneous Dirichlet boundary conditions.}
\end{table}

\vspace{.2cm}

In other words, by computing the spectrum of the
Laplacian on a disk of radius 1 (see table 1), the one given in
(\ref{eq:auto2d}) is the common eigenvalue of the  fourth and the
fifth eigenfunctions. Note that there are no eigenvalues with
multiplicity greater than 2. We also note that $\Phi =0$ at the boundary
of $\Omega$. These simple observations will be of primary
importance in the discussion to follow.

\begin{center}
\begin{figure}[h!]
\vspace{.6cm}
\centerline{\includegraphics[width=13.3cm,height=2.28cm]{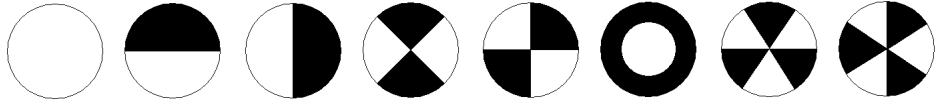}} 
\vspace{-.2cm}
\caption{\small Signature of the first 8 eigenfunctions of the Laplace's equation on a disk.
The first eigenfunction ($w_1$) does not change sign. The two successive ones ($w_2$ and $w_3$)
display a phase difference of 90 degrees. Then, we have $w_4$ and $w_5$ with
a phase difference of 45 degrees. The next one ($w_6$) is a single multiplicity
eigenfunction, while the last ones ($w_7$ and $w_8$) have a phase difference
of 30 degrees.}
\end{figure}
\end{center}

\begin{center}
\begin{figure}[h!]
\centerline{\includegraphics[width=13.3cm,height=2.28cm]{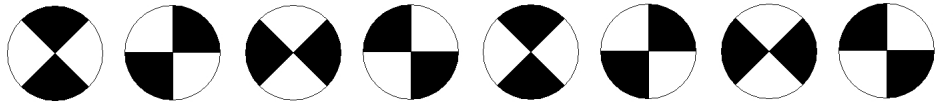}} 
\vspace{-.2cm}
\caption{\small Solutions of the wave equation on a disk are obtained by linearly combining
independent eigenfunction having the same eigenvalue. For example, a complete rotation 
is simulated through the sequence: $w_4, w_5, -w_4, -w_5, w_4, w_5, -w_4, -w_5$.
Intermediate situations are obtained by taking $w_4\cos ct\sqrt{\lambda} +w_5\sin ct
\sqrt{\lambda}$, with $0\leq ct\sqrt{\lambda}\leq 2\pi$, where $\lambda$ is the common eigenvalue.}
\end{figure}
\end{center}

\section{The Toroid Case}

We are ready to study the 3-D case.  In cylindrical coordinates
$(r,z,\phi)$, we consider the potentials:
\begin{equation}\label{eq:torpot}
{\bf A}~=~\frac {1} {c}\Big( a(t,r,z),~b(t,r,z),~0\Big)~~~~~~\Phi~=~-\left( \frac {\partial A}
{\partial r}+\frac {A} {r}+\frac {\partial B} {\partial z} \right)
\end{equation}
where $a$ and $b$ are functions to be computed, while $A$ and $B$
are their primitives with respect to the time variable. In this
context, the Lorenz condition (\ref{eq:color}) turns out to be
automatically satisfied. The functions $a$ and $b$ will describe
the time evolution, in a certain region $\Omega$ of  the plane
$(r,z)$, of a rotating wave similar to the one studied in the
previous section (differently from the previous case, $\Omega$ is
now vertically oriented). Since there is no dependance on $\phi$,
the solution is automatically extended to a toroid $\Sigma$,
having section $\Omega$, with the axis parallel to the $z$-axis.
\par\smallskip

From the potentials we deduce the electromagnetic fields (see (\ref{eq:potenz})):
\begin{equation}\label{eq:capneu}
{\bf E}~=~\Big( -\frac {\partial \Phi} {\partial r}-\frac {1} {c^2} \frac {\partial a} {\partial t},~
-\frac {\partial \Phi} {\partial z}-\frac {1} {c^2} \frac {\partial b} {\partial t},~0\Big)~~~~~~~~~
{\bf B}~=~\frac {1} {c^2}\Big(0,~0,~\frac {\partial b} {\partial r}-\frac {\partial a} {\partial
z}\Big)
\end{equation}
We now require ${\bf E}$ and ${\bf B}$ to satisfy the whole set of Maxwell equations (in alternative,
we get the same conclusions by imposing (\ref{eq:ondea}) and (\ref{eq:ondea1})). The result is the
following system:
\begin{equation}\label{eq:torsysa}
\frac {1} {c^2}\frac {\partial^2 a} {\partial t^2}~=~\frac{\partial^2 a}{\partial z^2} ~+~\frac
{\partial} {\partial r}\left(\frac {\partial a} {\partial r}~+~ \frac {a} {r}\right)
\end{equation}
\begin{equation}\label{eq:torsysb}
\frac {1} {c^2}\frac {\partial^2 b} {\partial t^2}~=~\frac{\partial^2 b}{\partial z^2} ~+~\frac
{\partial^2 b} {\partial r^2}~+~\frac {1} {r} \frac{\partial b} {\partial r}
\end{equation}
We can couple $a$ and $b$ through the boundary condition:
\begin{equation}\label{eq:torbc}
\frac {\partial a} {\partial z}~=~\frac {\partial b} {\partial r}
\end{equation}
which amounts to ask $~{\bf B}=c^{-1}{\rm curl}{\bf A}=0~$  at the contour of  $\Omega$. The field
${\bf E}$ is tangential to the same boundary. As a matter of fact, thanks to
(\ref{eq:torpot})-(\ref{eq:torsysa})-(\ref{eq:torsysb}), one can write:
\begin{equation}\label{eq:torsysbm}
{\bf E}~=~\left( \frac {\partial}{\partial z}\Big(\frac {\partial A}{\partial z} -\frac {\partial
B}{\partial r}\Big),~-\frac{1}{r} \frac {\partial}{\partial r} \Big( r \frac {\partial A}{\partial z}
-r\frac{\partial B}{\partial r}\Big) ,~0\right)
\end{equation}
which means  that ${\bf E}$ is orthogonal to the gradient  of $~r\Big(
\frac {\partial}{\partial z}A-\frac {\partial} {\partial r}B \Big)$.
Since (\ref{eq:torbc}) is valid for any $t$, such a gradient is
orthogonal to the boundary of $\Omega$, showing that  ${\bf E}$ is
tangential.
\par\smallskip

Note that now the  set $\Omega$ is not going to be a circle. As done
in Ref.~\cite{funaro}, section 5.4, we set $~y=r-\eta$. The domain
$\Omega$ will be centered at the point $(\eta ,0)$, where $\eta
>0$ is large enough to avoid  intersection of $\Omega$ with the
$z$-axis. The quantity $2\eta$ is related to the major diameter of
the toroid.
\par\smallskip

By differentiating the first equation with respect to $z$ and the
second one with respect to $y$, one gets:
\begin{equation}\label{eq:torpa1}
\begin{cases}\displaystyle \frac {1} {c^2} \frac {\partial^2 u}{\partial
t^2}~=~\frac {\partial^2 u}{\partial z^2} ~+~\frac {\partial} {\partial
y}\left(\frac {\partial u} {\partial y}+ \frac {u} {y+\eta}\right) \\ \\
\displaystyle \frac {1} {c^2} \frac {\partial^2 v}{\partial t^2}~=~\frac {\partial^2 v}{\partial z^2}
~+~\frac {\partial} {\partial y}\left(\frac {\partial v} {\partial y}+ \frac {v} {y+\eta}\right)
\end{cases}
\end{equation}
where we took  $u=\frac {\partial}{\partial z}a~$ and $~v=\frac {\partial}{\partial r}b =\frac
{\partial}{\partial y}b$, with $u-v=0~$ on $\partial\Omega$. Furthermore, with the substitutions
$~\tilde u=u\sqrt{y+\eta}$,
 $\tilde v=v\sqrt{y+\eta}$, one obtains:
\begin{equation}\label{eq:torpa2}
\begin{cases}\displaystyle \frac{1} {c^2}\frac {\partial^2 \tilde
u} {\partial t^2}~=~ \frac{\partial^2 \tilde u}{\partial z^2}~+~\frac {\partial^2 \tilde u}{\partial
y^2} ~-~\frac {3}{4}\frac {\tilde u} {(y+\eta)^2}
\\ \\ \displaystyle \frac{1} {c^2}\frac {\partial^2 \tilde
v} {\partial t^2}~=~ \frac{\partial^2 \tilde v}{\partial z^2}~+~\frac {\partial^2 \tilde v}{\partial
y^2} ~-~\frac {3}{4}\frac {\tilde v} {(y+\eta)^2}
\end{cases}
\end{equation}
to be solved in  $\Omega$,  with the boundary condition: $~\tilde u-\tilde v=0$.
\par\smallskip

Similarly to the case examined in section 2, we look for solutions
associated to the second mode ($k=2$  in (\ref{eq:cbdiskk})) with
respect to the angle of rotation. According to the captions 
of figures 1 and 2, the functions $\tilde u$ and
$\tilde v$ should have a phase difference of 45 degrees (this property is
now qualitative, since $\Omega$ is not a perfect disk). We fix
the area of $\Omega$ to be equal to $\pi$. For $\eta$ tending to
infinity, the set $\Omega$ converges to a circle of radius 1 and
the electromagnetic fields coincide with those given in
(\ref{eq:cbdisk}) for $\omega =\delta_0$.
\par\smallskip

We proceed by introducing a new  unknown $w=\tilde u-\tilde v$ and
by taking the difference of the two equations in
(\ref{eq:torpa2}). Then, we can get rid of the time variable and
pass to the stationary eigenvalue problem:
\begin{equation}\label{eq:torpa3}
\begin{cases}\displaystyle Lw~=~-~\frac {\partial^2 w}{\partial z^2}~-~
\frac {\partial^2 w}{\partial y^2}~+~\frac {3} {4}\frac {w}{(y+\eta)^2}~= ~\lambda w~~~~{\rm
in}~\Omega \\  \displaystyle w~=~0~~~~~~~{\rm on}~\partial\Omega \end{cases}
\end{equation}
Thus, we would like $\lambda >0$ to be  an eigenvalue of the
positive-definite differential operator: $L=-\frac {\partial^2}
{\partial z^2}- \frac {\partial^2} {\partial y^2}+\frac {3}
{4}(y+\eta )^{-2}$, with  homogeneous Dirichlet boundary
conditions. In addition, since we want two independent
eigenfunctions with a difference of phase of 45 degrees, the
multiplicity of $\lambda$ must be equal to 2. This implicitly
defines $\Omega$. More precisely, we require that $\lambda
=\lambda_4=\lambda_5$, where $\lambda_4$ and $\lambda_5$ are the
eigenvalues of the fourth  and the fifth eigenfunctions, $w_4$ and
$w_5$, of $L$ on $\Omega$. In order to preserve energy, these
eigenfunctions will be normalized in $L^2(\Omega )$. In this way,
by setting:
\begin{equation}\label{eq:solon}
\tilde u (t,y,z)-\tilde v (t,y,z)~=~w_4(y,z)\sin (ct\sqrt{\lambda })~+~w_5(y,z)\cos (ct\sqrt{\lambda
})
\end{equation}
one gets a full solution of the system (\ref{eq:torpa2}). We can finally recover the electromagnetic
fields by the expressions (see (\ref{eq:torsysbm})):
$$
{\bf E}=\frac {1} {c\sqrt{\lambda}}\left( \frac {\partial} {\partial z}\Big(\frac {-w_4\cos \zeta
+w_5\sin \zeta}{\sqrt{y+\eta}}\Big),  ~~~~~~~~~~~~~~  \right.
$$
$$  \left. ~~~~~~~~~~~~~~~~~~
~ \frac{-1}{\sqrt{y+\eta}}\frac {\partial} {\partial y}\Big( (-w_4\cos \zeta
+w_5\sin \zeta )\sqrt{y+\eta}\Big),~0\right)$$
\begin{equation}\label{eq:camp}
{\bf B}~=~\frac {1}{c^2}\left( 0,~~0,~~\frac {w_4\sin \zeta +w_5\cos \zeta} {\sqrt{y+\eta}}\right)
\end{equation}
with $\zeta =ct\sqrt{\lambda }$. As in figure 2, by varying the parameter $t$, we
can simulate a rotating object. 
\par\smallskip

Of course, we could also take
into consideration the case $k>2$ (see the 2-D analog
(\ref{eq:cbdiskk})) by studying the behavior of eigenvalues with
higher magnitude. The search, in this case, should be addressed
to the determination of couples of independent eigenfunctions
sharing the same eigenvalue. We think this is a viable option,
although, in order to maintain the discussion at a simple
level, we will not analyze this extension.
\par\smallskip

We now define $\Sigma =\Omega\times [0,2\pi [$,  in order to get a
3-D solution not depending on $\phi$. In fluid dynamics this
structure is known as {\sl vortex ring}  (see for instance Ref.~\cite
{batchelor}, section 7.2, or ~\cite{lim}). On the  surface of
$\Sigma$, field  ${\bf E}$ is tangential and ${\bf B}$ is zero. At
every point inside $\Sigma$, the time average during a period of
oscillation, of both ${\bf E}$ and ${\bf B}$,  is zero (see the
animations in  Ref.~\cite{funweb}).
\par\smallskip

We would like to know more about  the shape of $\Sigma$.
Considering that not all the sets $\Omega$ are such that
$\lambda_4$ and $\lambda_5$ are equal, we can use this property in
order to determine the right domain. This problem admits however
infinite solutions. In the next section we discuss how to find
numerically some suitable configurations.

\section{On the Optimal Shape of $\Omega$}

We first observe that,  among the sets with fixed area equal to
$\pi$, the circle of radius 1, minimizes the eigenvalues of the
Laplacian with homogeneous Dirichlet boundary conditions (see for
instance Ref.~\cite{henrot}). Due to symmetry arguments, all the
eigenvalues related to the angular modes have multiplicity 2 (see
table 1 and figure 1). The eigenfunctions are supposed to be orthogonal and
normalized in $L^2(\Omega) $. Then, the normal derivative of the
eigenfunction related to the lowest eigenvalue, has constant value
on the boundary of the disk.  Similarly, the sum of the squares of
the normal derivatives of the eigenfunctions related to the second
and the third eigenvalues, is constant along the boundary. The
same is true for all the couples of eigenfunctions relative to
other eigenvalues with double multiplicity (the fourth and the
fifth, for example).
\par\smallskip

We now replace the Laplacian  by the new elliptic operator
$L=-\frac {\partial^2} {\partial z^2}- \frac {\partial^2}
{\partial y^2}+\frac {3} {4}(y+\eta )^{-2}$ (see
(\ref{eq:torpa3})), with homogeneous Dirichlet boundary
conditions. We are concerned with finding a set $\Omega$, with
area equal to $\pi$, such that the fourth and the fifth
eigenvalues,  $\lambda_4$ and $\lambda_5$, of $L$ are coincident.
Thanks to (\ref{eq:solon}), this allows us to determine solutions
of a wave-type equation, rotating inside $\Omega$ with an angular
velocity proportional to $\sqrt{\lambda}$, where $\lambda
=\lambda_4=\lambda_5$. In table 2, we report the eigenvalues of
$L$ on the disk  centered in $(y,z)=(0,0)$ and radius equal to 1.
If $~0<\eta <1~$ the differential operator is singular inside
$\Omega$ (the corresponding toroid region $\Sigma$ has no central
hole). Therefore, we take $\eta \geq 1$. The case $\eta =1$, where
the axis $y=-\eta$ (or, equivalently, $r=0$) touches $\Omega$ at
the boundary, is still admissible. The table shows the results for
different values of $\eta$.  For $\eta$ large, the fourth and the
fifth eigenvalues are quite similar, due to the fact that the term
$~\frac {3} {4}(y+\eta )^{-2}~$ becomes small. For $\eta$ close to
1, the situation is not too bad, anyway. This is especially true
for the highest eigenvalues, since the corresponding
eigenfunctions, due to the boundary constraints, decay faster near
the border of $\Omega$. Therefore, they do not ``feel'' too much the
presence of the term $~\frac {3} {4}(y+\eta )^{-2}$.

\vspace{.5cm}
\begin{table}[ht!]
\label{tab2} \noindent\[
\begin{array}{|c|c|c|c|c|}
  \hline
   &   ~~~~\eta=1 ~~~~~  &    ~~~~\eta=1.2 ~~~~  &    ~~~~\eta=1.5 ~~~~   &    ~~~~\eta=2 ~~~~~ \\
  \hline
  \lambda_1 & ~~ 6.90 & ~~ 6.46 & ~~6.18 & ~~5.99  \\
  \lambda_2 & 15.71 & 15.33 & 15.08 & 14.90  \\
  \lambda_3 & 16.75 & 15.68 & 15.19 & 14.93  \\
  \lambda_4 & 28.06 & 27.27 & 26.86 & 26.63  \\
  \lambda_5 & 28.36 & 27.35 & 26.88 & 26.64  \\
    \hline
\end{array}
\]
\caption{\small The first 5 eigenvalues of the operator $L$ on a disk of radius 1, for different values of
$\eta$.}
\end{table}
\par\smallskip

For any fixed $\eta$, we would  like to adjust the shape of
$\Omega$ with the help of some iterative procedure. The idea is to
correct the boundary in order to fulfill a certain condition at
the limit. Before getting some interesting answers (see later), we
tried unsuccessfully different approaches. Let us discuss the one
that looked more promising. We consider the substitution:
\begin{equation}\label{eq:subs}
\hat w(y,z)~=~\frac {w(s,z)}{\sqrt{y+\eta }}~~~~~{\rm where}~~s=(y+\eta )^2
\end{equation}
In this way (\ref{eq:torpa3}) becomes:
\begin{equation}\label{eq:torpa5}
\begin{cases}\displaystyle -\Big(4s~\frac {\partial^2 \hat w}{\partial s^2}~+~
\frac {\partial^2 \hat w}{\partial z^2}\Big)~=~\lambda \hat w~~~~{\rm in}~\Omega \\  \\ \displaystyle
\hat w~=~0~~~~~~~{\rm on}~\partial\Omega
\end{cases}
\end{equation}
where now the operator is a Laplacian with a variable coefficient.
Concerning the operator $L_\beta =-\beta \frac
{\partial^2}{\partial s^2}-\frac {\partial^2}{\partial z^2}$, for
$\beta >0$ constant, the domain $\Omega$ of fixed area equal to
$\pi$, optimizing the set of eigenvalues, is an ellipse. As in the
case of the circle ($\beta =1$), we have infinite eigenvalues with
double multiplicity. In addition, if $\hat w$ is the first
eigenfunction of $L_\beta$, the following relation is easily
checked:
\begin{equation}\label{eq:torpa6b}
\beta \Big(\frac {\partial \hat w}{\partial s}\Big)^2~+~ \Big(\frac {\partial \hat w}{\partial
z}\Big)^2~=~{\rm constant}~{\rm on}~\partial\Omega
\end{equation}
The idea is to generalize the above relation in the case of variable coefficients. Hence, let us
suppose that $\hat w$ is an eigenfunction corresponding to the first eigenvalue in (\ref{eq:torpa5}),
we may require that:
\begin{equation}\label{eq:torpa6}
4s \Big(\frac {\partial \hat w}{\partial s}\Big)^2~+~ \Big(\frac {\partial \hat w}{\partial
z}\Big)^2~=~{\rm constant}~{\rm on}~\partial\Omega
\end{equation}
For points where $s$ is large, we expect the curvature of the boundary of $\Omega$ to be high, and
viceversa. The resulting $\Sigma$ is a kind of doughnut, a bit flattened on the internal side. Due to
(\ref{eq:subs}), in terms of $w$ one has:
$$(y+\eta ) \left[\Big(\frac {w}{2\sqrt{y+\eta}}+\frac {\partial w} {\partial y}\Big)^2+
\Big(\frac{\partial w}{\partial z}\Big)^2\right]$$
\begin{equation}\label{eq:torpa7}
=~(y+\eta ) \left[\Big(\frac {\partial w}{\partial
y}\Big)^2+ \Big(\frac{\partial w}{\partial z}\Big)^2\right]~ =~ {\rm constant}
\end{equation}
where we used that $w=0$ on $\partial \Omega$.
\par\smallskip

Relation (\ref{eq:torpa7}) is the ``target''  we would like to reach
by implementing our iterative method. To this end, starting from
the circle of radius 1, we compute the eigenfunction (normalized
in $L^2(\Omega )$) corresponding to the lowest eigenvalue in
(\ref{eq:torpa3}). We then compute its normal derivatives on
$\partial\Omega$ and correct $\Omega$ in order to enforce
condition (\ref{eq:torpa7}). For points belonging to the boundary,
the updating procedure is performed as follows:
\begin{equation}\label{eq:itera}
{\bf P}_{\it new}~=~{\bf P}_{\it old}~+~\theta {\cal E}{\bf N}
\end{equation}
where ${\bf N}$ is the outer normal derivative,  $\theta$ is a
relaxation parameter and ${\cal E}$ is the difference between the
normal derivative evaluated in ${\bf P}_{\it old}$ and the average
of the normal derivatives computed on the boundary of the current
$\Omega$. After every correction, the area of $\Omega$ is set to
$\pi$, with the help of a linear transformation. At the limit, we
would like to have ${\cal E}=0$.
\par\smallskip

In order to compute eigenvalues and eigenfunctions we  used a
finite element code. In particular, we implemented MODULEF (see
Ref.~\cite{modulef} and ~\cite{george}) with $P_2$ elements. The grid
was fine enough to have reliable results up to the second decimal
digit (at least for the first eigenvalue). The  solver is based on
a QR algorithm. The points to be updated through the iterative
method (\ref{eq:itera}) are the vertices of the triangles
belonging to $\partial \Omega$. The normal derivative at these
points is the average of the normal derivatives at the mid-points
of the two sides of the contiguous triangles. After the entire
boundary has been modified, a brand-new grid is generated in
$\Omega$. The performances of this technique are far from being
optimal, but their improvement is not in the scopes of the present
paper.
\par\smallskip

Unfortunately, despite all the efforts, the method did not  want
to converge. A possible explanation is that relation
(\ref{eq:torpa7}) cannot be realized, because it is wrong from the
theoretical viewpoint. The iterative technique (\ref{eq:itera})
was however a good starting point to try successive variants, and,
finally, we had success with the following scheme:
\begin{equation}\label{eq:itera2}
{\bf P}_{\it new}~=~{\bf P}_{\it old}~+~\theta ~\vert \lambda_4-\lambda_5\vert~\Big( n^{(y)},~~
(y_{\it old}+\eta )n^{(z)}\Big)~~~~~{\rm on}~\partial\Omega
\end{equation}
where ${\bf N}=(n^{(y)},n^{(z)})$ is again the outer  normal
derivative. This time, the fourth and the fifth eigenvalues of $L$
explicitly appear in the correcting term. Inspired by
(\ref{eq:torpa7}), the updating on the boundary is not uniform in
the variable $y$, but the second component is weighted by the
function $y+\eta $.
\par\smallskip

After suitably adjusting $\theta$,  convergence is obtained  in a
few iterations. The result is a domain $\Omega$ yielding
$\lambda_4=\lambda_5$. However, as anticipated, there are infinite
other domains with this property. The most interesting ones should
be those having a distribution of the eigenvalues as lower as
possible, and making the shape of $\Omega$ as rounded as possible.
We do not know if what we got by iterating (\ref{eq:itera2}) can
be considered optimal with this respect. Nevertheless, it is
important to have shown that the condition $\lambda_4=\lambda_5$
can be actually achieved (at least numerically). In table 3, one
finds the modified eigenvalues for various $\eta$ (compare with
table 2). Note instead that, for $\eta$ close to 1, $\lambda_2$
and $\lambda_3$ are not in good agreement.
\par\smallskip

For $\eta =1$ and $\eta =1.2$, the corresponding  domains are
displayed in figure 3. Even for $\eta$ close to 1, they are not
too far from a circle. In figure 4, for $\eta =1$, we provide the
plot of the fourth and the fifth eigenfunctions, actually showing 
a phase difference of 45 degrees (see also figures 1 and 2). For $\eta\geq 2$,
$\Omega$ is practically a circle.
\par\smallskip

The analysis carried out in  this section may help to understand
the structure of fluid dynamics vortex rings. These are incredibly
stable configurations, that have been widely studied, under
different aspects. Among the vast literature we just mention for
instance the papers: ~\cite{maxworthy}, \cite{pullin},
\cite{shariff}, \cite{wakelin}. In addition, in Ref.~\cite{linden},
``optimal'' sections of vortex rings are discussed. However, in the
fluid dynamics case, the situations where the major diameter is
relatively small, bring to more pronounced deformations of the
sections.

\vspace{.5cm}
\begin{table}[ht]
\label{tab3} \noindent\[
\begin{array}{|c|c|c|c|c|}
  \hline
   &   ~~~~\eta=1 ~~~~~  &    ~~~\eta=1.2 ~~~~~  &    ~~~\eta=1.5 ~~~~~   &    ~~~~\eta=2 ~~~~~ \\
  \hline
\lambda_1  &~~6.85 & ~~6.44 & ~~6.17 & 5.98 \\
  \lambda_2  & 15.76 & 15.32 & 15.04 &  14.89\\
  \lambda_3  & 16.59 & 15.64 & 15.18 & 14.90  \\
  \lambda_4  & 28.08 & 27.24 & 26.83 & 26.60 \\
  \lambda_5  & 28.08 & 27.24 & 26.83 & 26.60 \\
    \hline
\end{array}
\]
\caption{\small The first 5 eigenvalues of the operator $L$ on the modified domain, for different values of
$\eta$.}
\end{table}
\par\smallskip

We are not able to discuss the  stability of our electromagnetic
vortices. Theoretically,  using the newfound 3-D solutions, one
should build the corresponding electromagnetic stress tensor and
put it on the right-hand side of Einstein's equation, as
illustrated in Ref.~\cite{funaro}. Then, one has to solve this
nonlinear system in order to determine the metric tensor. Finally,
one should check that the quasi-circular orbits ($\Omega$ is not a
perfect circle) followed by the light rays are actually geodesics
of such a space-time environment. Note that this study depends on
the parameter $\omega$, connected to the speed of rotation of the
wave. For $\omega$ large, the size of $\Omega$ is  small and the
frequency is high. We expect one or more situations of
equilibrium, where the gravitational setting and the centrifugal
effect of the spinning  solitons are compensated. This would mean
that only particular geometries are admitted, specifying exactly
the shape and the size of the toroid regions.

\begin{center}
\begin{figure}[h!]
\centerline{\includegraphics[width=4.5cm,height=5.3cm]{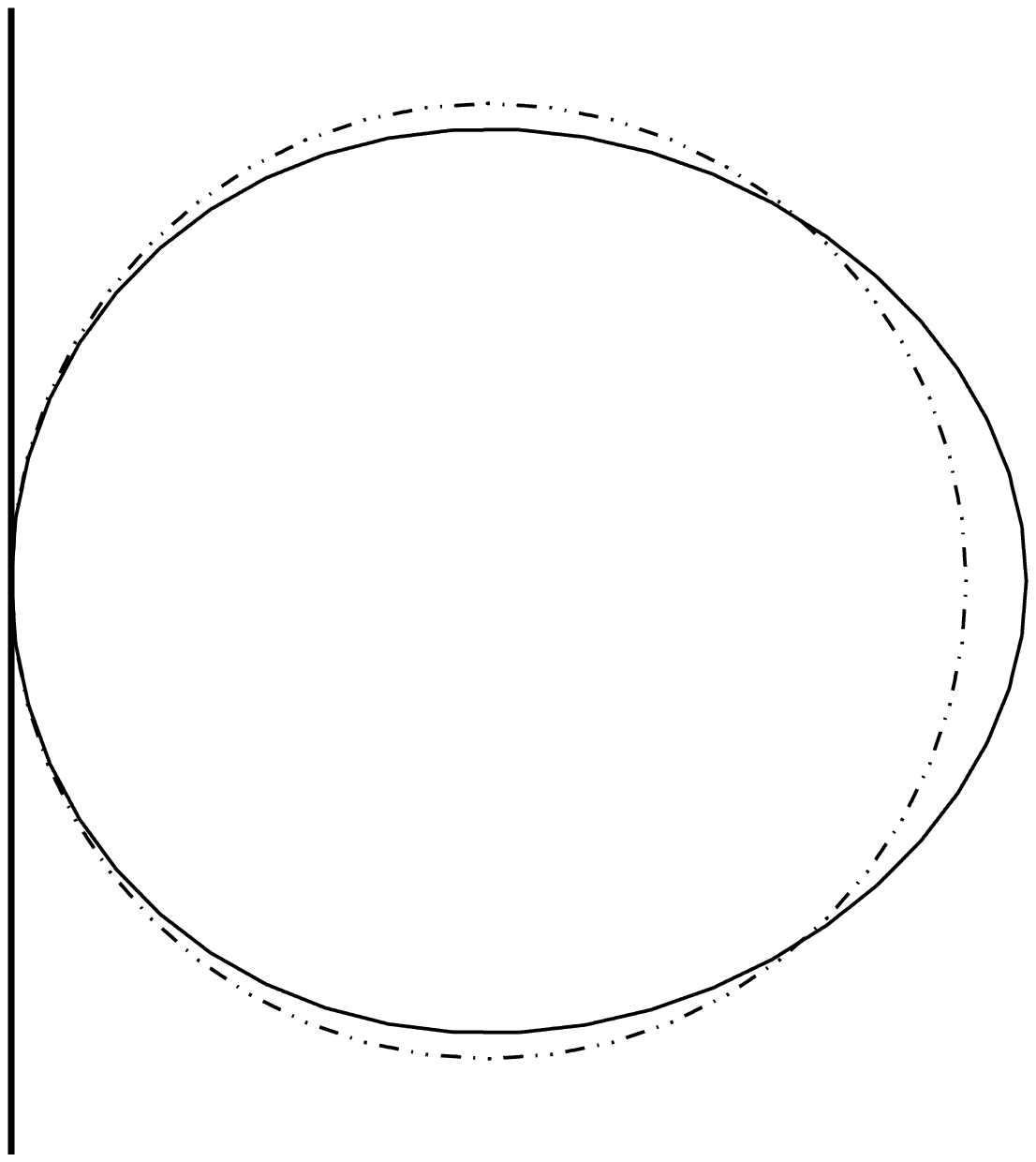}\hspace{1.2cm}
\includegraphics[width=5.2cm,height=5.3cm]{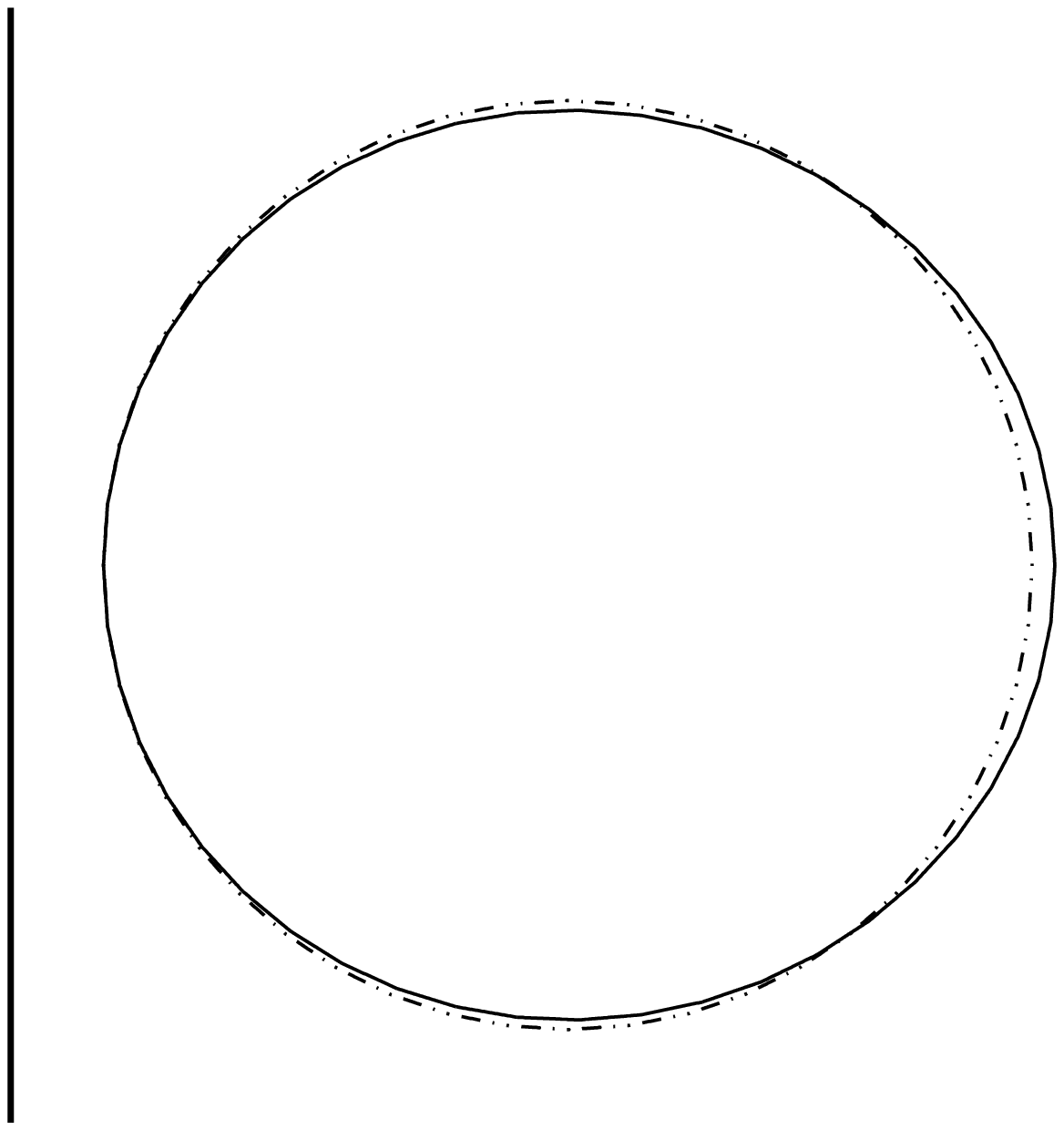}} 
\vspace{-.1cm}
\vspace*{8pt}
\caption{\small  Toroid sections (solid line) compared to the circle of radius 1
(dashed-dotted line), for  $\eta =1$ and $\eta =1.2$. The vertical line is the toroid axis of
symmetry.}
\vspace{-.2cm}
\end{figure}
\end{center}

\begin{center}
\begin{figure}[!h]
\centerline{\includegraphics[width=5.cm,height=5.cm]{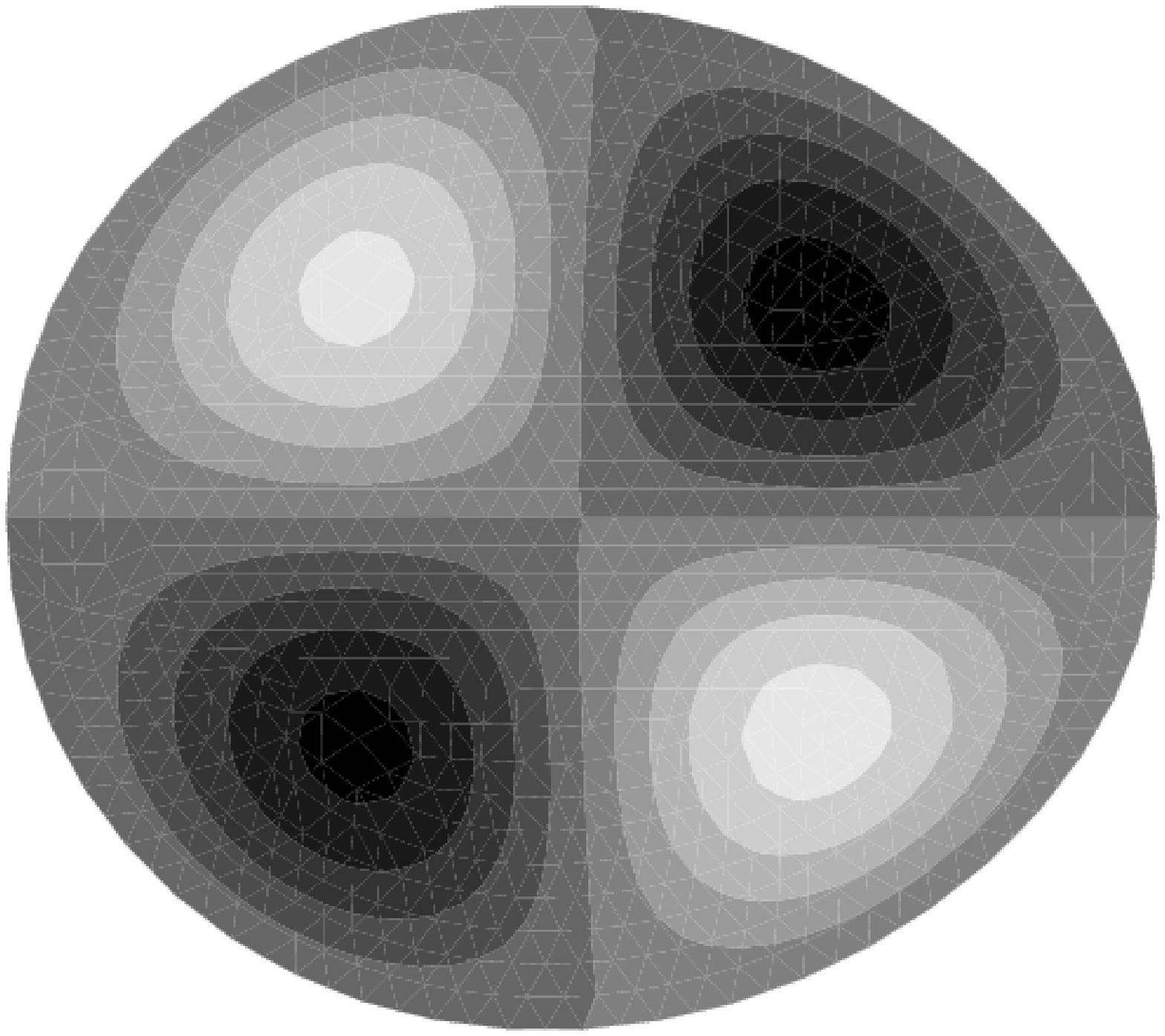}\hspace{1.1cm}
\includegraphics[width=5.cm,height=5.cm]{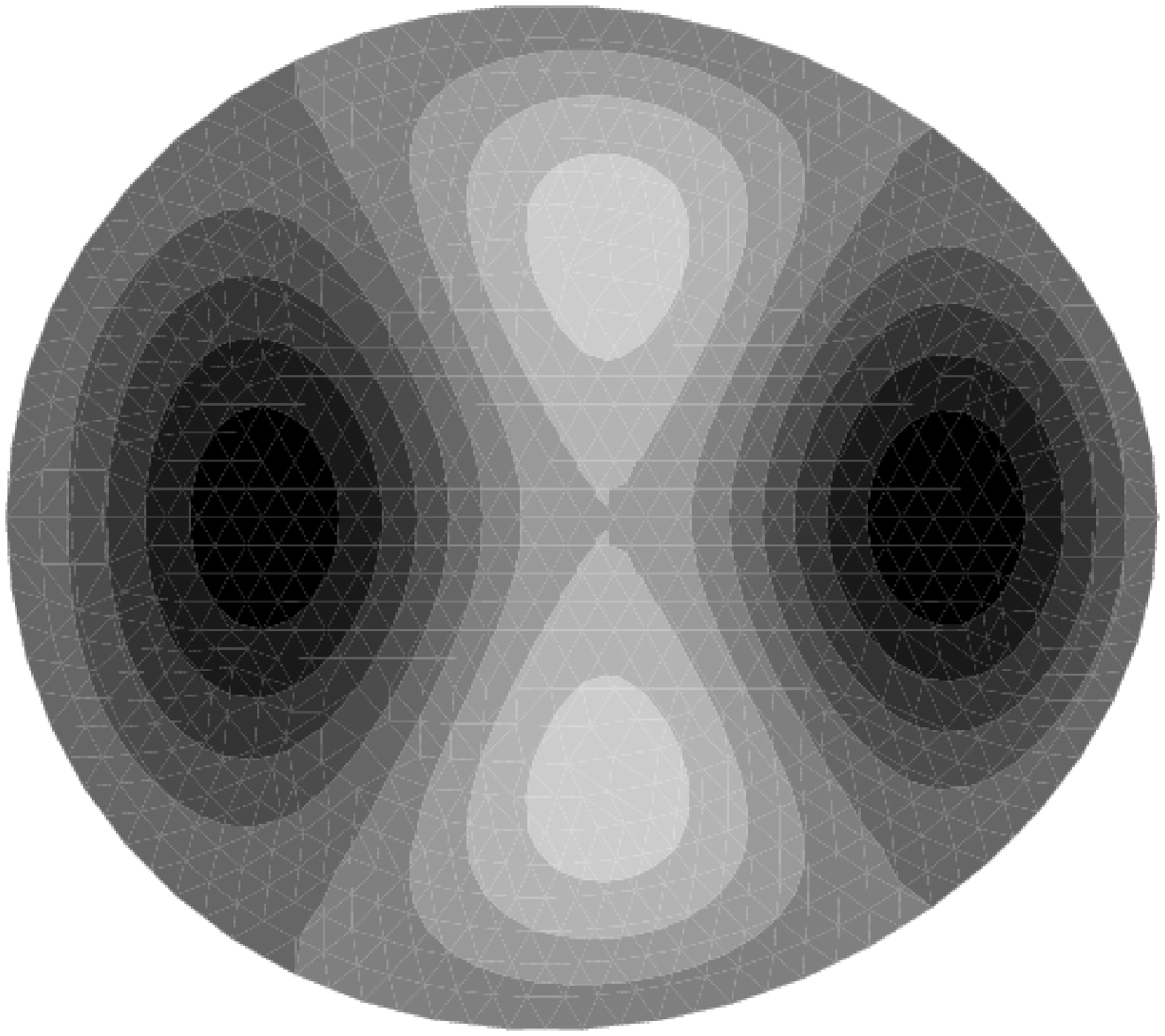}} 
\vspace{-.1cm}
\vspace*{8pt}
\caption {\small The forth and the fifth normalized eigenfunctions corresponding to
the case $\eta =1$. Even if $\Omega$ is not a perfect circle, they actually display a phase
difference of 45 degrees. Each eigenfunction has two positive (white) and two negative (black)
bumps.}
\end{figure}
\end{center}

\par\smallskip

Of course, what we just said above  turns out to be very hard to
prove, both theoretically and computationally. Note that we are
also omitting a stationary component of the electromagnetic
fields, introduced in Ref.~\cite{funaro}, which should provide charge
and mass to the particle model. We do not insist further on this
topic and we address the reader to Ref.~\cite{funaro} for additional
information.

\section{Hill's Type Vortices}

The next step is to examine what happens outside the spinning
toroid.  If we were dealing with a fluid vortex, due to viscosity,
we might expect the formation of other external vortices
developing at lower frequency. This circumstance can be observed
in tornados or typhoons (see for instance Ref.~\cite{kuo}). In the
theory presented in Ref.~\cite{funaro}, an electromagnetic vortex,
through a mechanism still to be clarified, captures the
surrounding electromagnetic signals and generates a series of
encapsulated shells vibrating with decreasing frequencies. These
shells could be responsible for the quantum properties of matter.

\begin{center}
\begin{figure}[!h]
\centerline{\includegraphics[width=4.9cm,height=5.4cm]{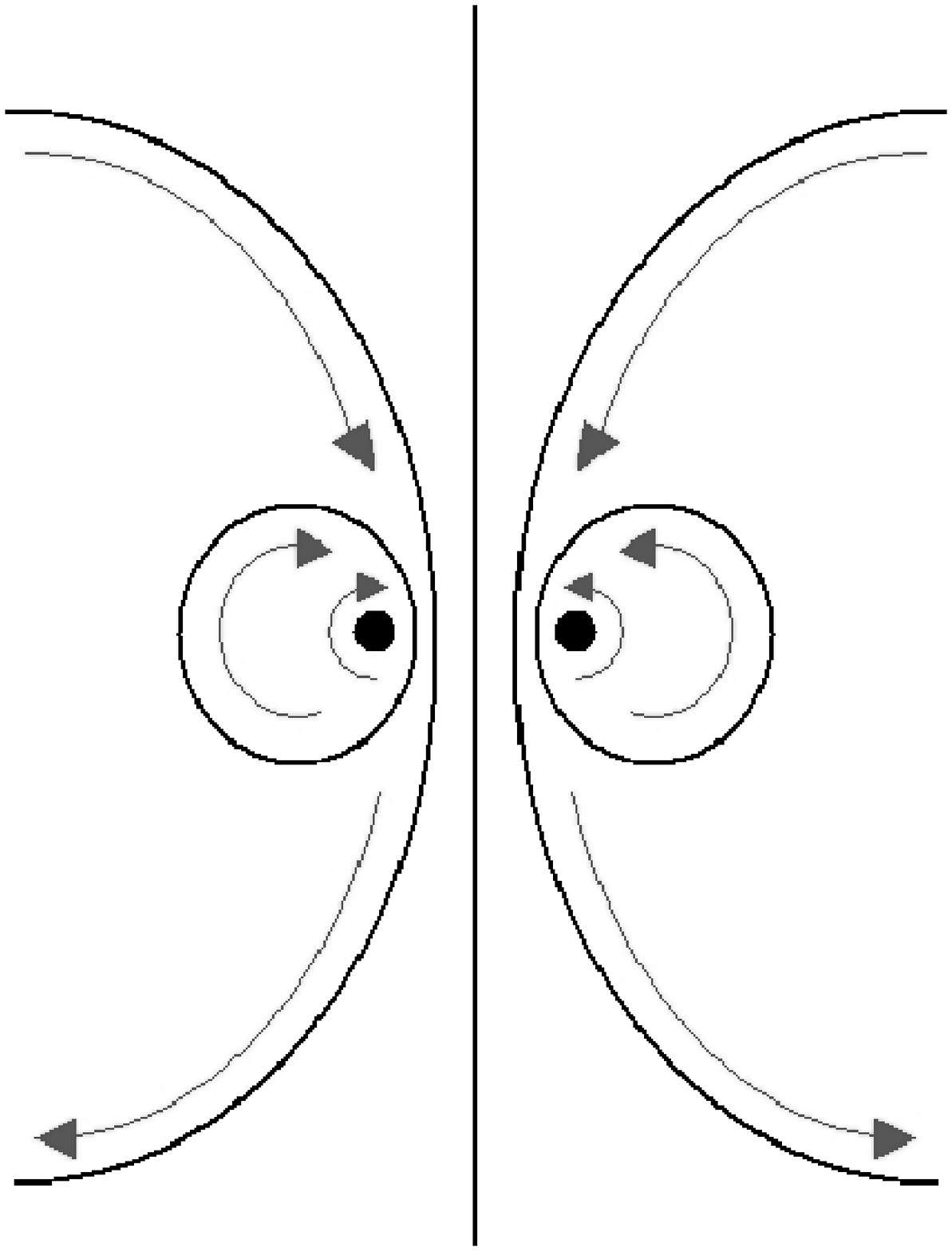}\hspace{1.2cm}
\includegraphics[width=4.9cm,height=5.4cm]{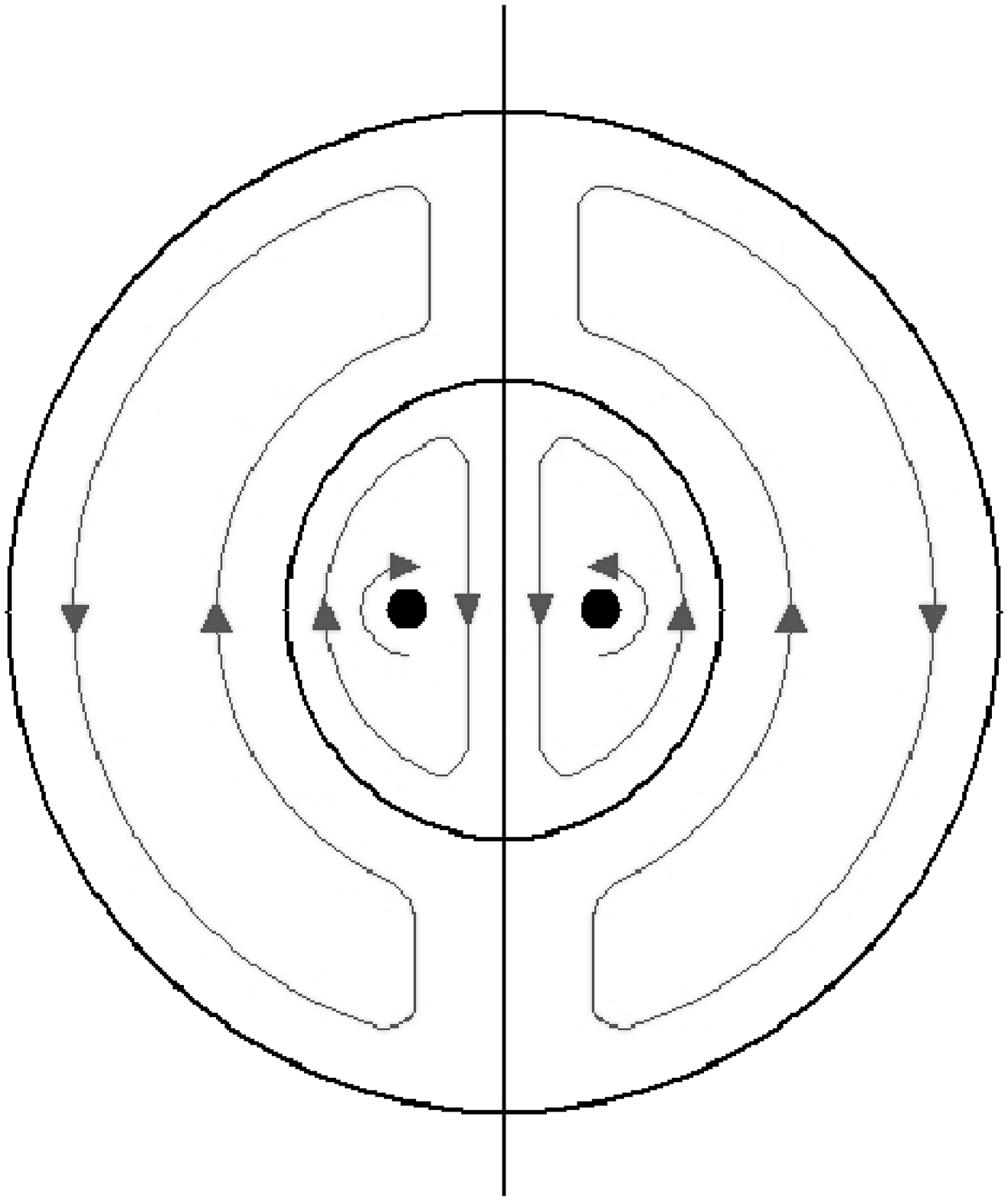}} \vspace{.1cm}
\caption{\small Successive layers of encapsulated vortex rings. In the first
picture, the toroid regions have a common central hole. In the second one, the central hole is
reduced to a vertical segment (Hill's vortex). Of course, intermediate situations combining some
shells of the first type, successively embedded in spherical vortices, could be taken into
consideration.}
\end{figure}
\end{center}

Basically, there are two possible ways  in which external ring
vortices may develop. As the first picture of figure 5 shows, we
may have a series of successive toroid structures, where all the
fluid stream-lines pass through a common central hole. This
situation is difficult to analyze, since we have no idea of the
shape of these regions and the location of the primary vortex
inside them. In practice, there are too many degrees of freedom to
work with.
\par\smallskip

The other situation (second picture of figure 5) is more
affordable. It represents an Hill's spherical vortex (see
Ref.~\cite{batchelor}, section 7.2, and Ref.~\cite{elcrat} for some
computational results), successively surrounded by other spherical
layers. Thus, we now know exactly the shape of these structures.
We just have to find the location (and the relative size) of the
primary toroid vortex inside the most internal sphere. This
research will lead us to interesting conclusions.
\par\smallskip

\begin{center}
\begin{figure}[ht]
\centerline{\includegraphics[width=13.cm,height=6.5cm]{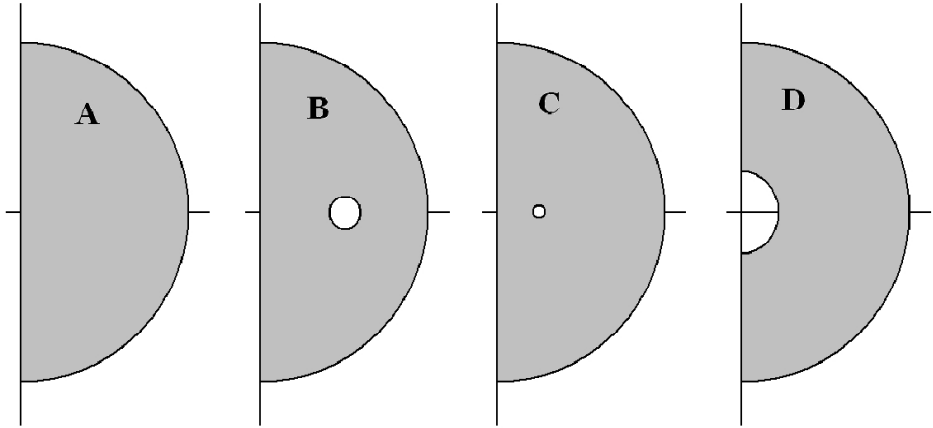}}\vspace{-.3cm}
\vspace*{8pt}
\caption{\small Sections of spherical vortices.}
\end{figure}
\end{center}\vspace{-.5cm}

We start by considering the domain $A$  in figure 6. We set the
radius equal to 1, so that the area is $\pi /2$. We solve in
$\Omega =A$ the eigenvalue problem (\ref{eq:torpa3}), involving
the operator $L=-\frac {\partial^2}{\partial
z^2}-\frac{\partial^2}{\partial r^2} +\frac{3} {4r^2}$ ($r=y+\eta$
with $\eta =0$). In the first column of table 4, we report the
approximated eigenvalues, obtained by discretization with the
finite element method. As the reader can notice, the fourth and
the fifth eigenvalues are not coincident. Therefore, for a perfect
spherical vortex, we have no chances to obtain solutions of the
time-dependent wave equation. In alternative, we could modify a
bit the shape of the domain $A$ or accept solutions whose
stream-lines are not stationary.  This is not however the path we
would like to follow. If the rotation inside $\Omega$ is generated
by an internal spinning toroid, we may cut out a hole (as in the
domains $B$ and $C$) and see what happens to the spectrum of $L$.
For simplicity, the hole will be a circle of a certain radius,
suitably placed in a specific spot (although we know from section
4 that such a hole is slightly deformed). Playing with the size
and the location of the circle, we look for situations in which
the fourth and the fifth  eigenvalues ($\lambda_4$ and
$\lambda_5$) of $L$ are the same.
\par\smallskip

Surprisingly, this research seems to have a finite  number of
solutions. According to table 4 (columns 2 and 3), we have
coincidence of $\lambda_4$ and $\lambda_5$ in two particular
cases. In the first one, the center of the small circle is placed
at point $(\frac{1} {2},0)$ and the radius is approximately equal
to $r_B=.0505$. In the second one, the center is at  $(\frac {1}
{4},0)$ and the radius is approximately equal to $r_C=.00520$.
With the exception of these cases, by varying the magnitude and
the position of the internal circles, we always found
$\lambda_4\not = \lambda_5$. In our experiments we did  not try
anyway all the possible configurations, thus we do not exclude the
existence of other significant settings.

\begin{table}[!h]\vspace{.5cm}
\label{tab4} \noindent\[
\begin{array}{|c|c|c|c|c|}
  \hline
   &   ~{domain}~A ~  &   ~{domain}~B  ~  &    ~{domain}~ C~   &   ~{domain}~D ~ \\
  \hline
  \lambda_1 & 20.24 & 30.23 & 21.64 & 0.63 \\
  \lambda_2 & 33.29 & 34.16 & 33.32 & 0.99 \\
  \lambda_3 & 48.94 & 54.82 & 49.09 & 1.45 \\
  \lambda_4 & 59.82 & 68.15 & 67.17 & 1.99 \\
  \lambda_5 & 67.11 & 68.15 & 67.17 & 1.99 \\
    \hline
\end{array}
\]
\caption{\small The first 5 eigenvalues of the operator $L$, based on the four domains given in figure 6,
with homogeneous Dirichlet boundary conditions.}
\end{table}

We provide in figure 7 the discretization grids.  We show in
figures 8 and 9, the time evolution of the rotating waves,
obtained from expression (\ref{eq:solon}) for equispaced values of
$t$ with $c\sqrt{\lambda}t\in [0,\pi]$ (half-cycle). In the first
case, the spherical flow is chained to the inner toroid. In the
second case, the flow avoids the toroid, still remaining inside
the spherical region. The sequence looks rather complicated. With
the help of some imagination, one can see two anti-clockwise
rotating waves, circulating independently in the lower and the
upper quarters. They have different phases, so that the positive
(white) bumps and the negative (black) ones, alternately merge to
form a single protuberance situated near the center. The pictures
only show half of the cycle. Then, the sequence restarts with the
two colors interchanged. The corresponding animations can be found in
Ref.~\cite{funweb} (click {\it related papers}).
\par\smallskip

We point out once again that we are dealing with electromagnetic
waves.  Therefore, we should ask ourselves what happens to the
vector fields. Our plots actually show the evolution of the
function $~\frac{\partial}{\partial r}b -\frac{\partial}{\partial
z}a$, where ${\bf A}=c^{-1}(a,b,0)$ is the vector potential. Using
(\ref{eq:capneu}), (\ref{eq:torsysbm}) and (\ref{eq:camp}), one
then computes ${\bf E}$ and ${\bf B}$. We discover that, at each
point, when time passes, the tips of the arrows of the electric
field turn around, describing elliptic orbits (see the animations
in Ref.~\cite{funweb} in the case of a circle). The average of ${\bf
E}$ during a cycle is zero. The frequency of rotation is globally
the same, but, depending on the point, it is associated with a
different phase, so that the general framework seems quite
unorganized.

\begin{center}
\begin{figure}[!h]
\vspace{.2cm}
\centerline{\includegraphics[width=4.8cm,height=8.cm]{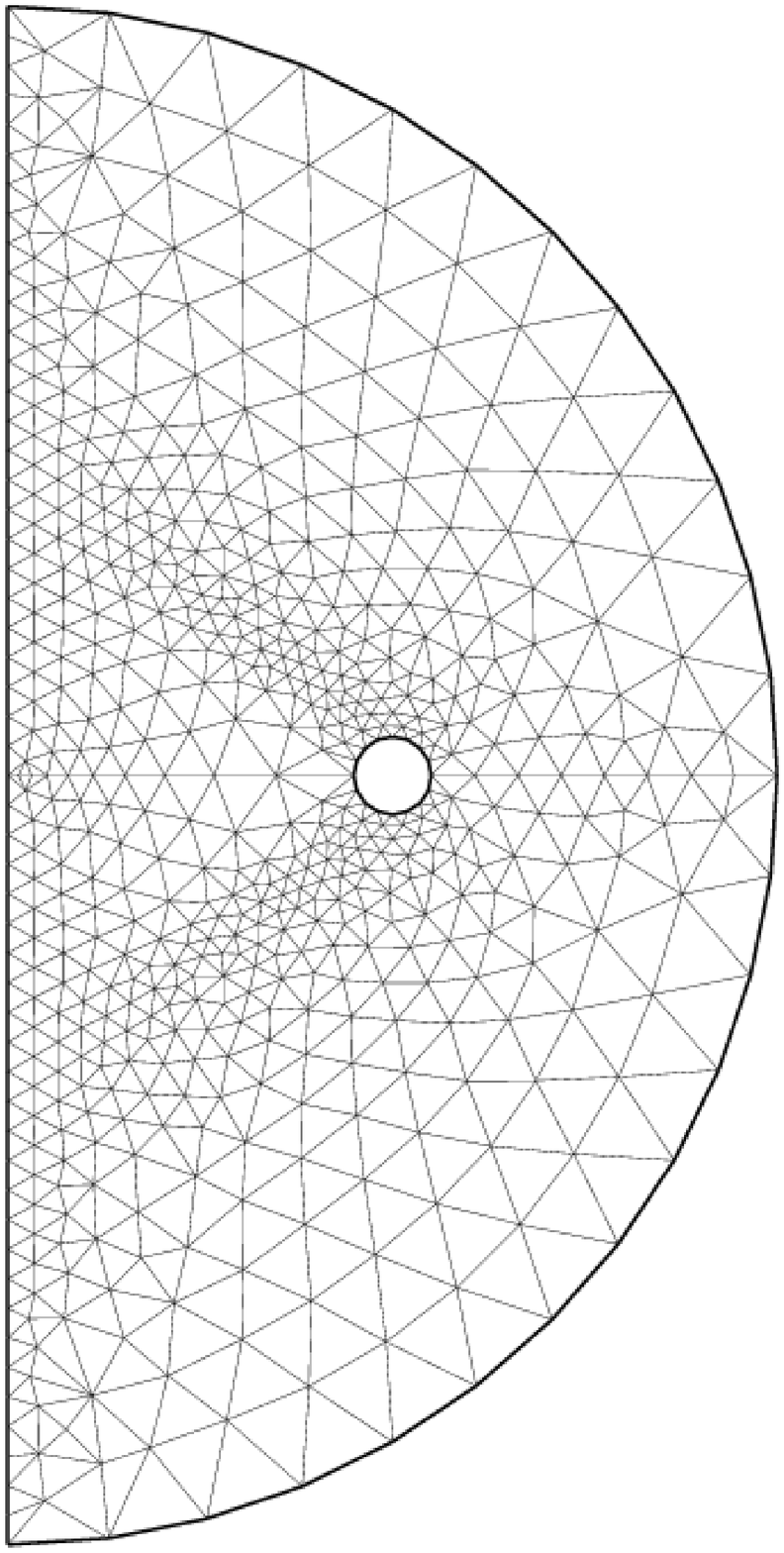}
\hspace{1.5cm}\includegraphics[width=4.8cm,height=8.cm]{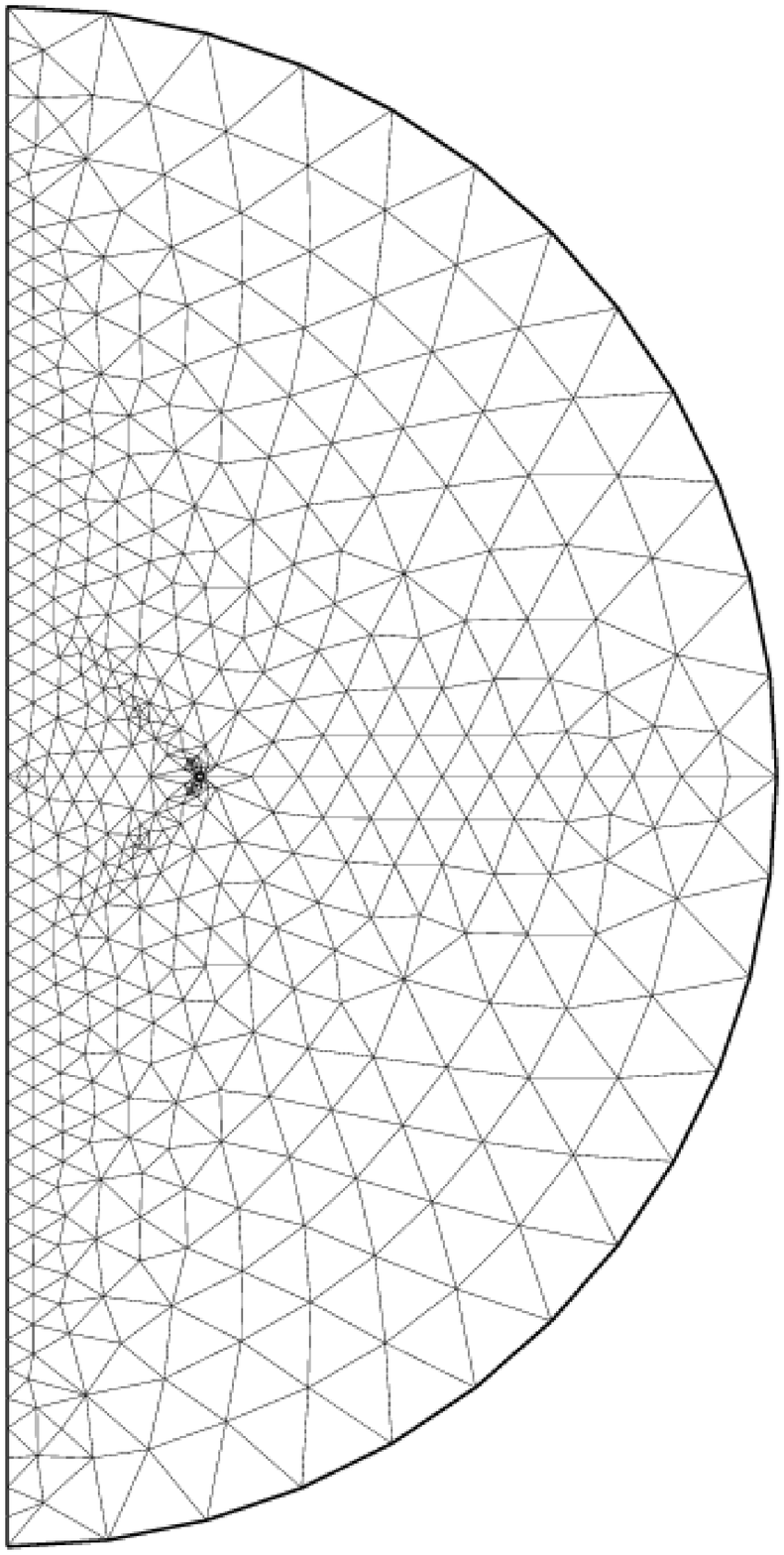} }
\caption{\small Discretization grids for the domains $B$ and $C$. The mesh is finer
near the vertical axis since the differential operator $L$ is singular there. The degrees of freedom
(internal and boundary nodes) are 2324 and 1968, respectively. In the right-hand side picture, the
hole is very small.}
\end{figure}
\end{center}

\begin{center}
\begin{figure}[!ph]
\centerline{\includegraphics[width=3.1cm,height=4.4cm]{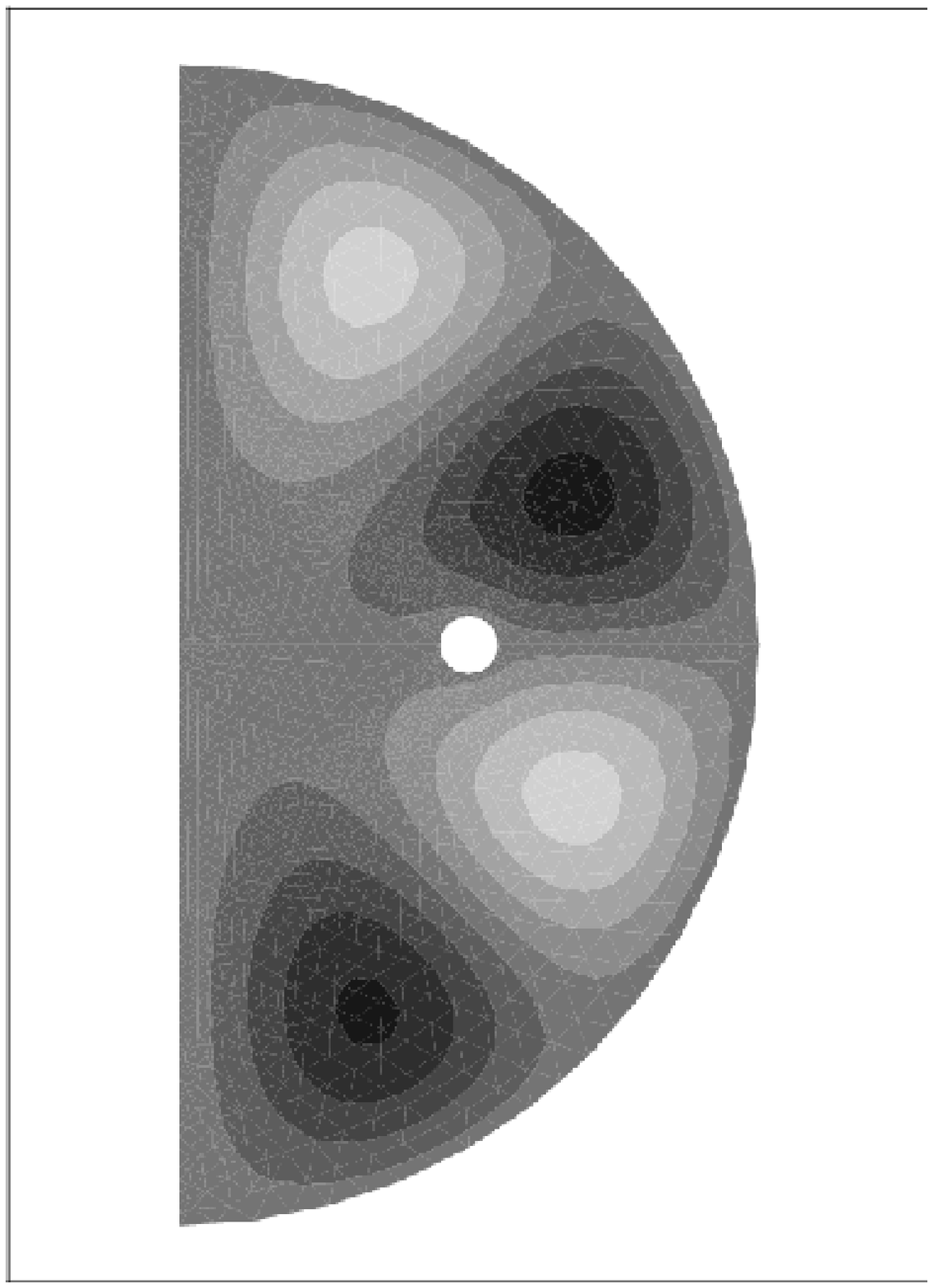}
\hspace{.06cm}\includegraphics[width=3.1cm,height=4.4cm]{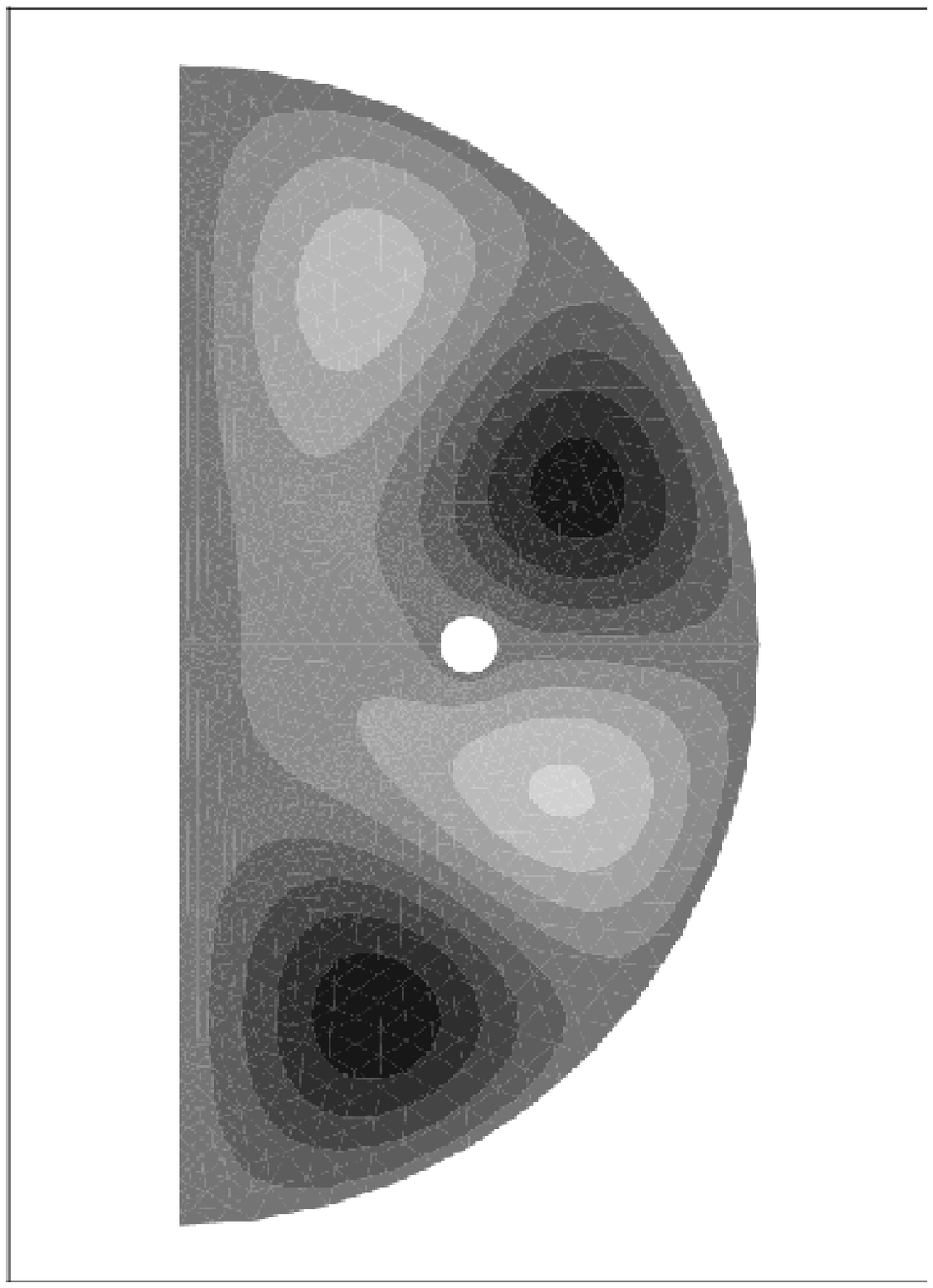}
\hspace{.06cm}\includegraphics[width=3.1cm,height=4.4cm]{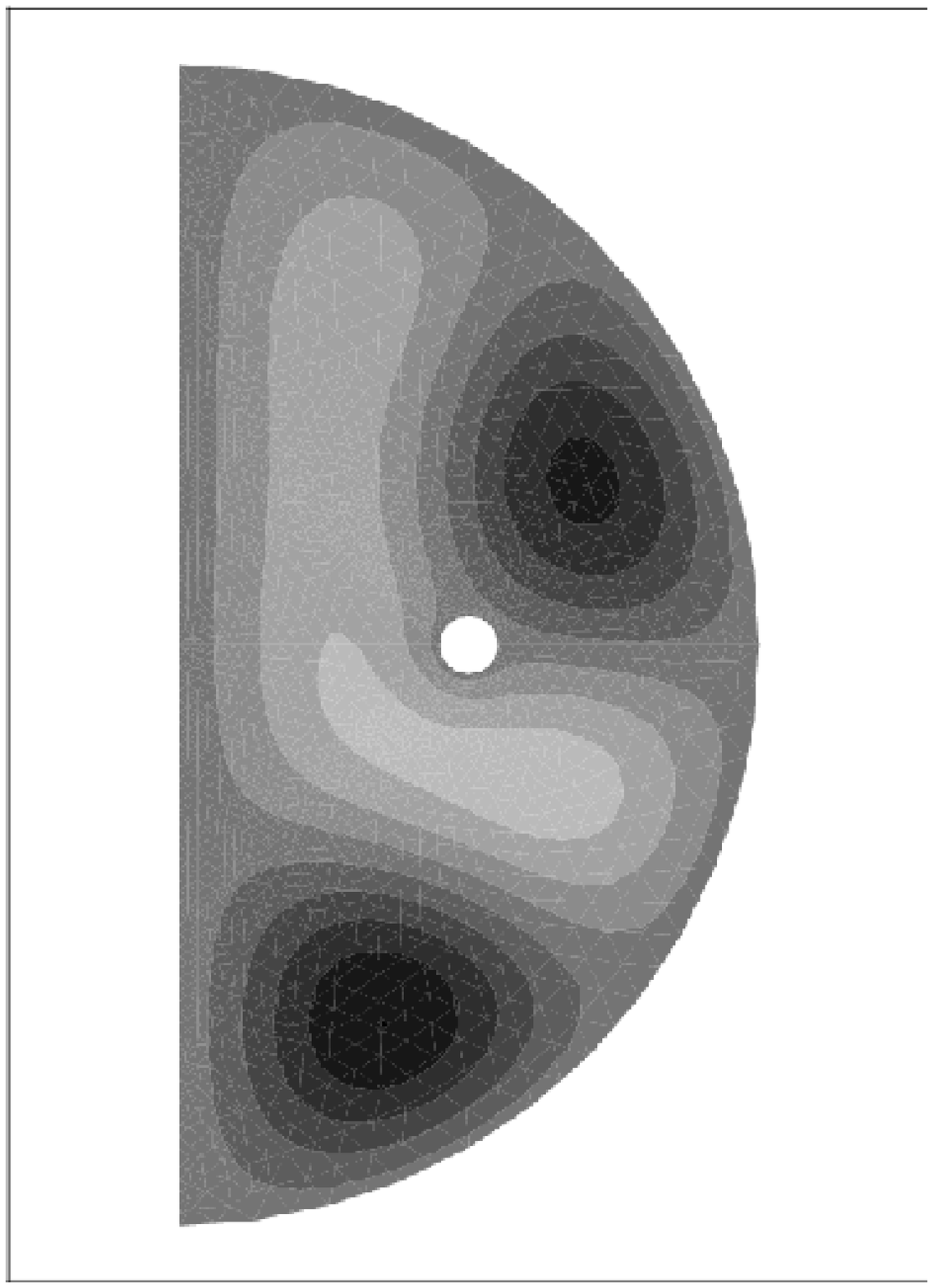}
\hspace{.06cm}\includegraphics[width=3.1cm,height=4.4cm]{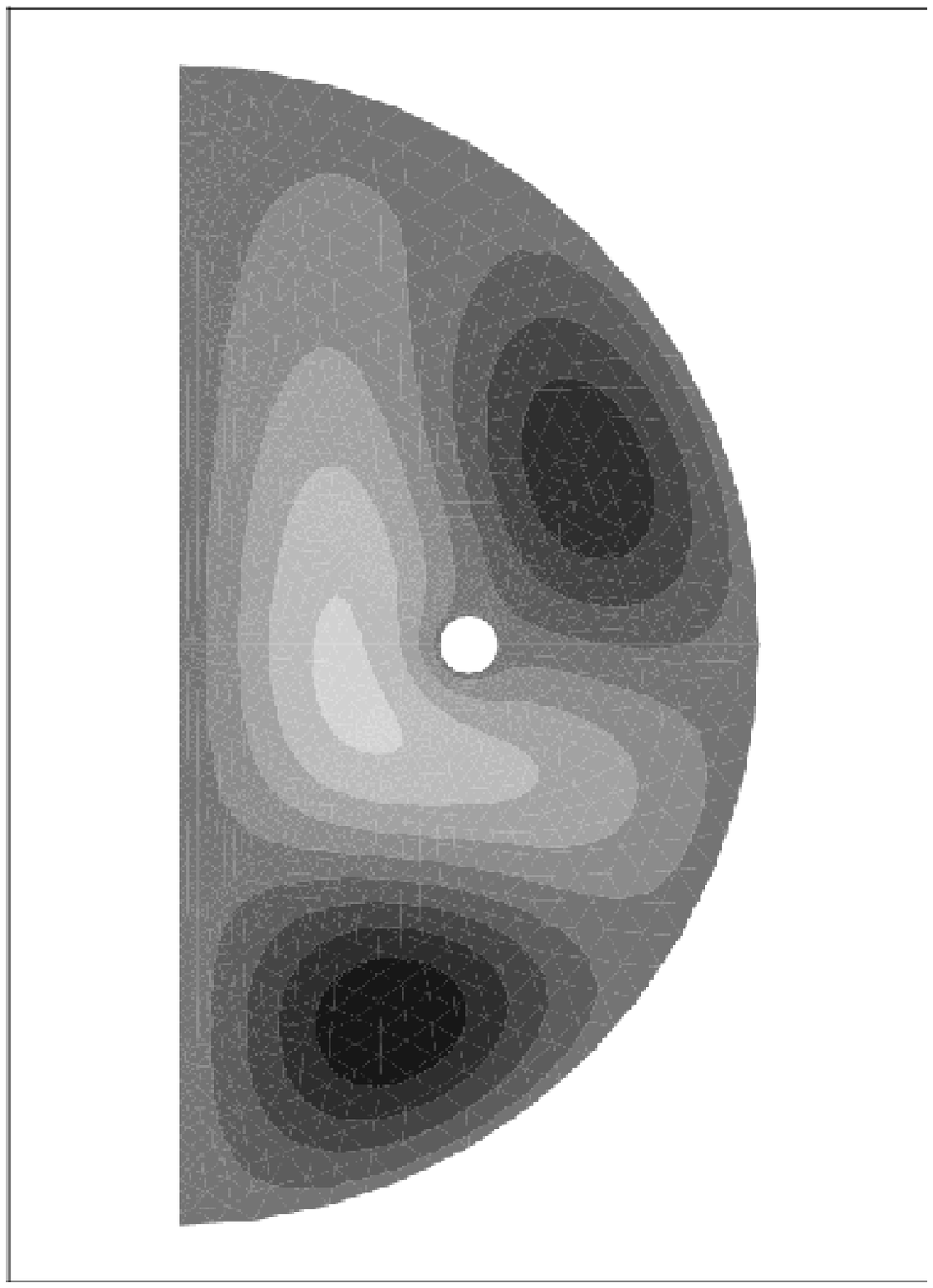}} \vspace{.07cm}
\centerline{\includegraphics[width=3.1cm,height=4.4cm]{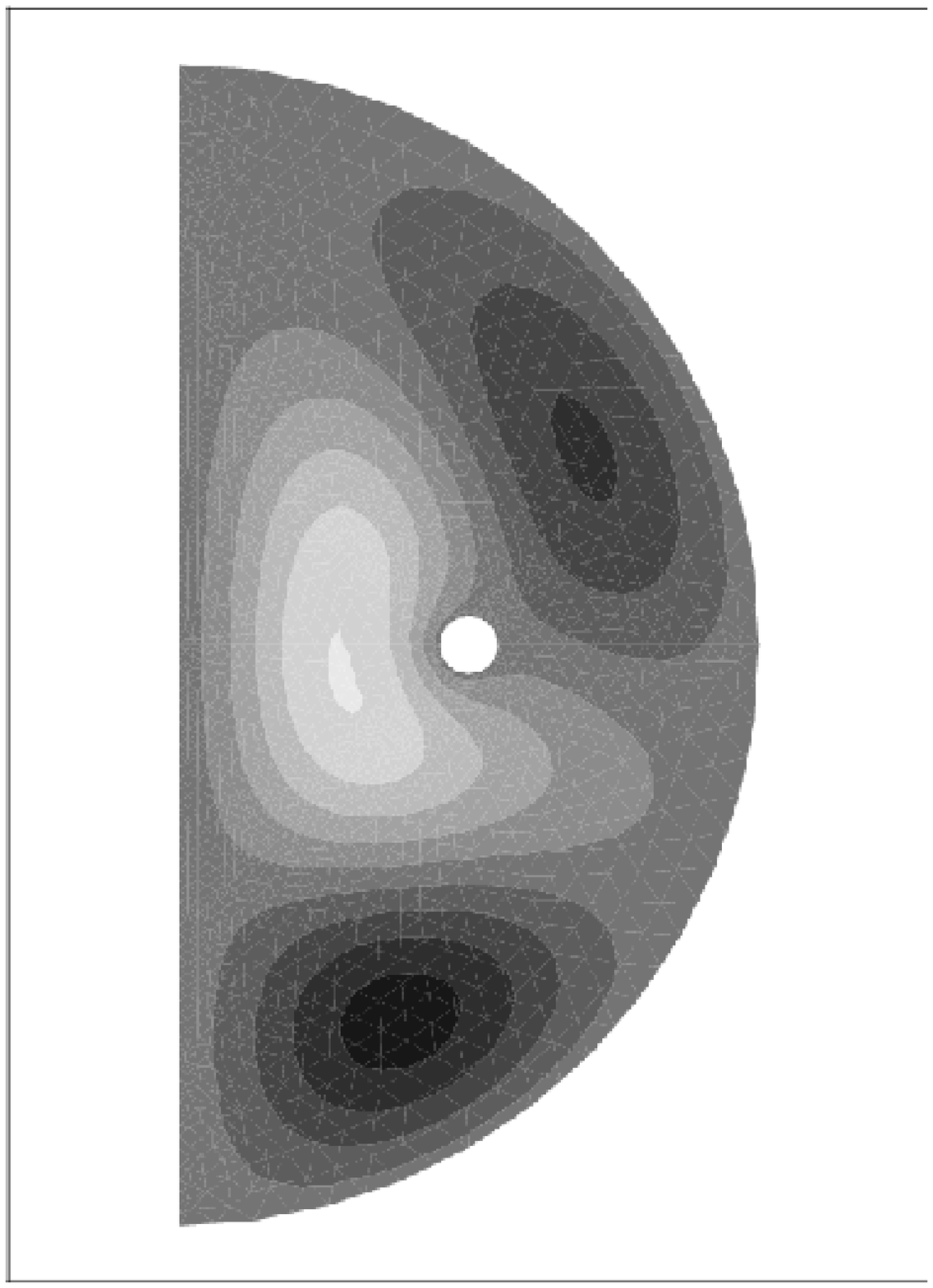}
\hspace{.06cm}\includegraphics[width=3.1cm,height=4.4cm]{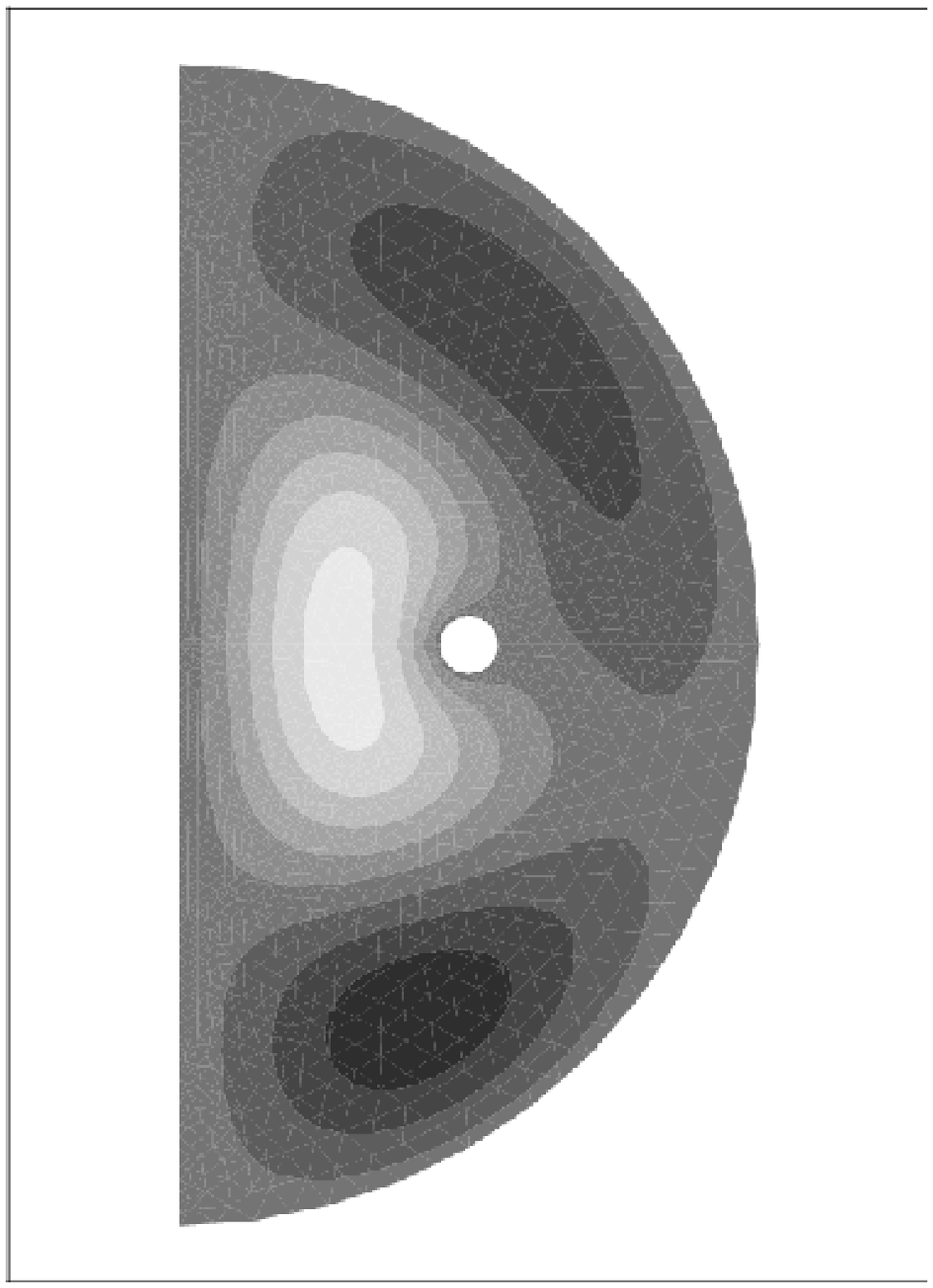}
\hspace{.06cm}\includegraphics[width=3.1cm,height=4.4cm]{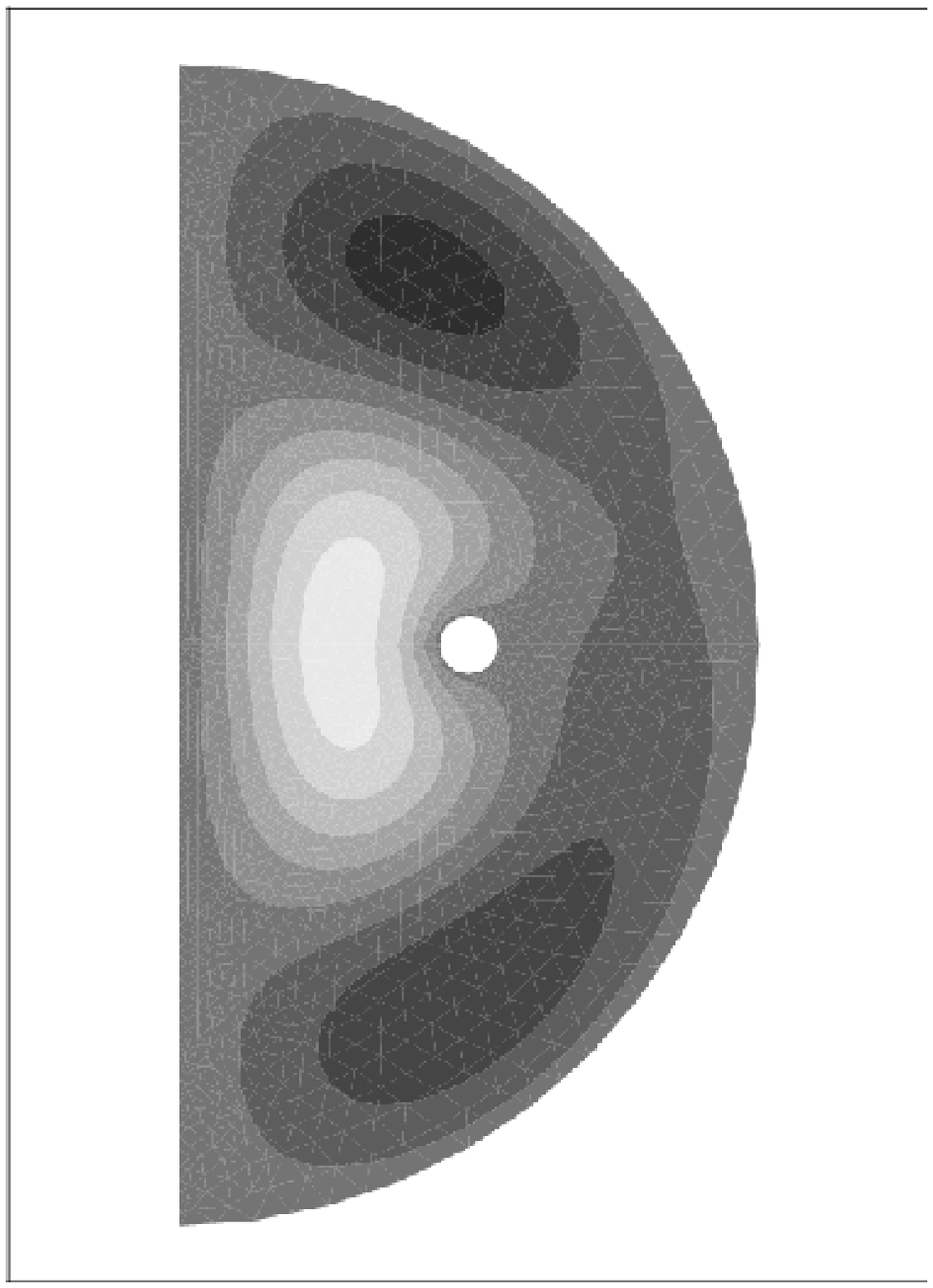}
\hspace{.06cm}\includegraphics[width=3.1cm,height=4.4cm]{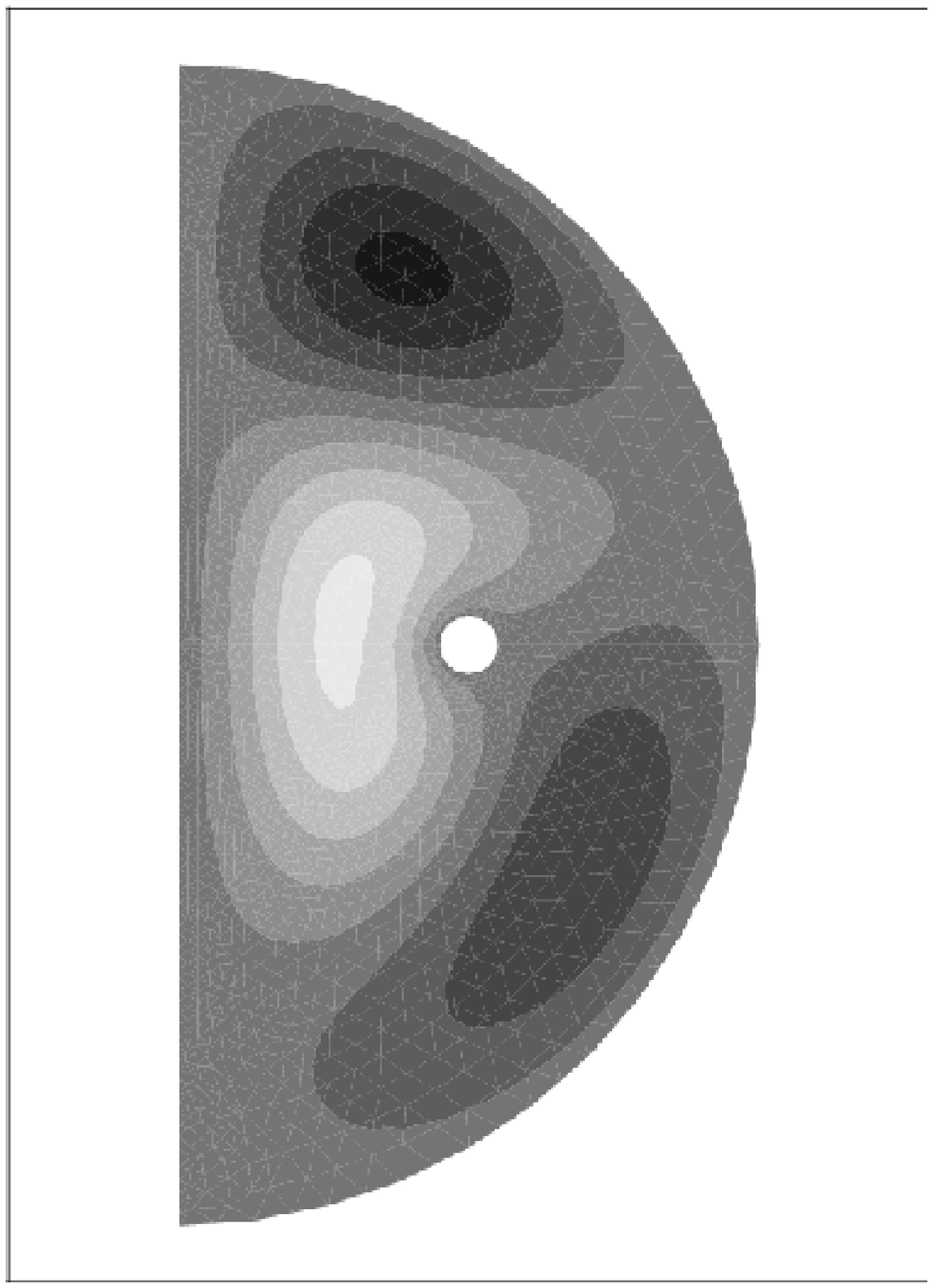}} \vspace{.07cm}
\centerline{\includegraphics[width=3.1cm,height=4.4cm]{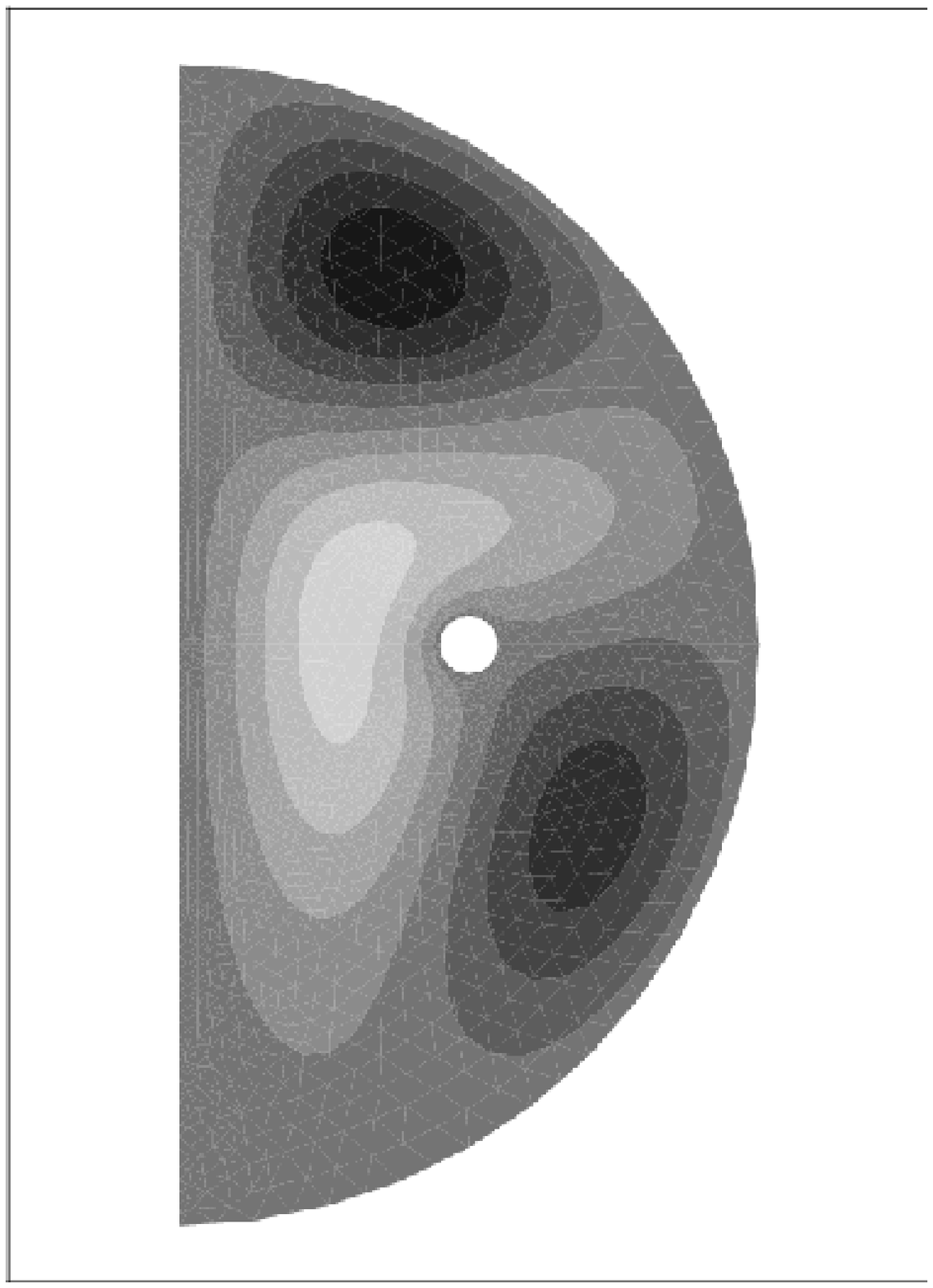}
\hspace{.06cm}\includegraphics[width=3.1cm,height=4.4cm]{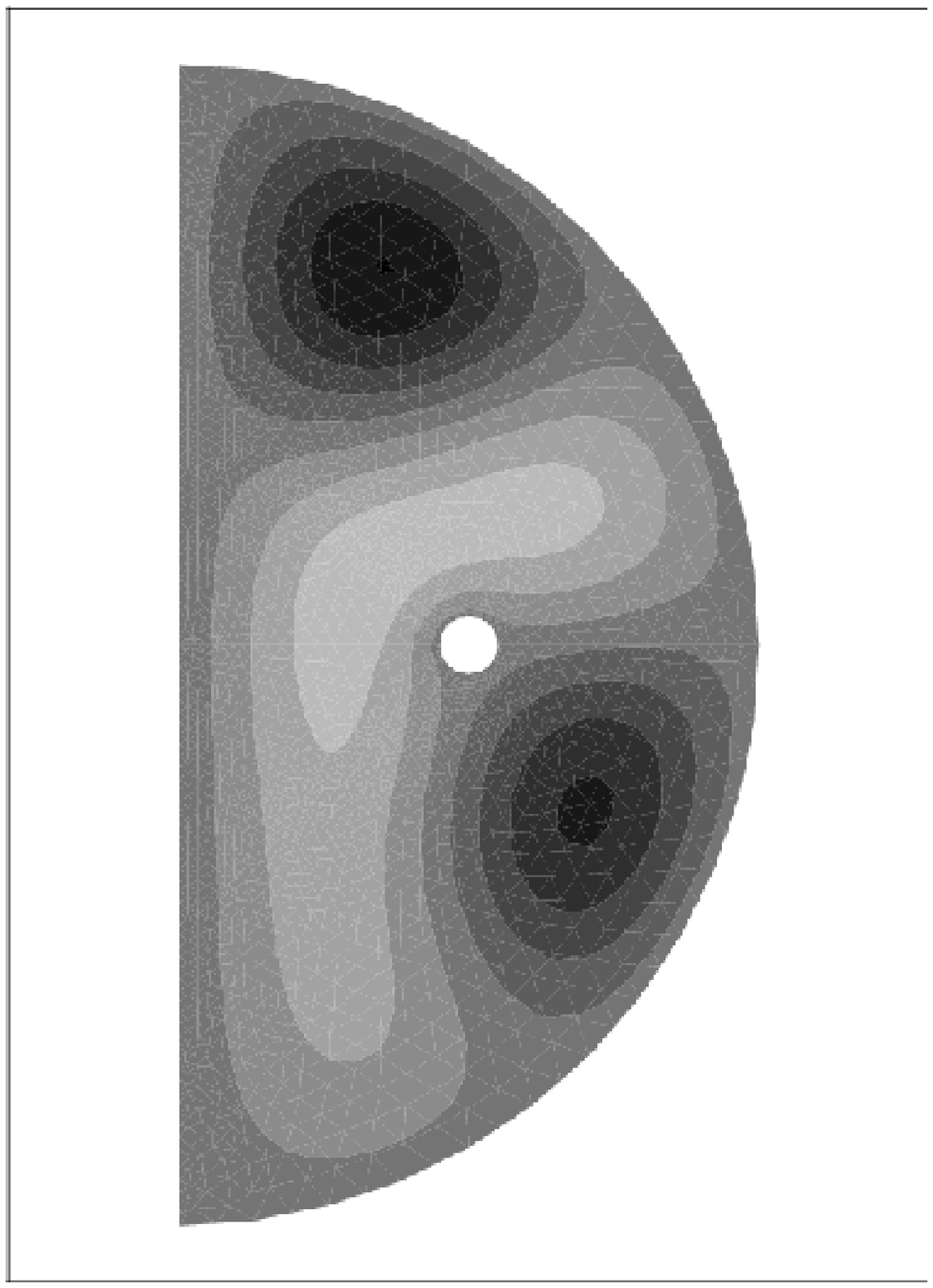}
\hspace{.06cm}\includegraphics[width=3.1cm,height=4.4cm]{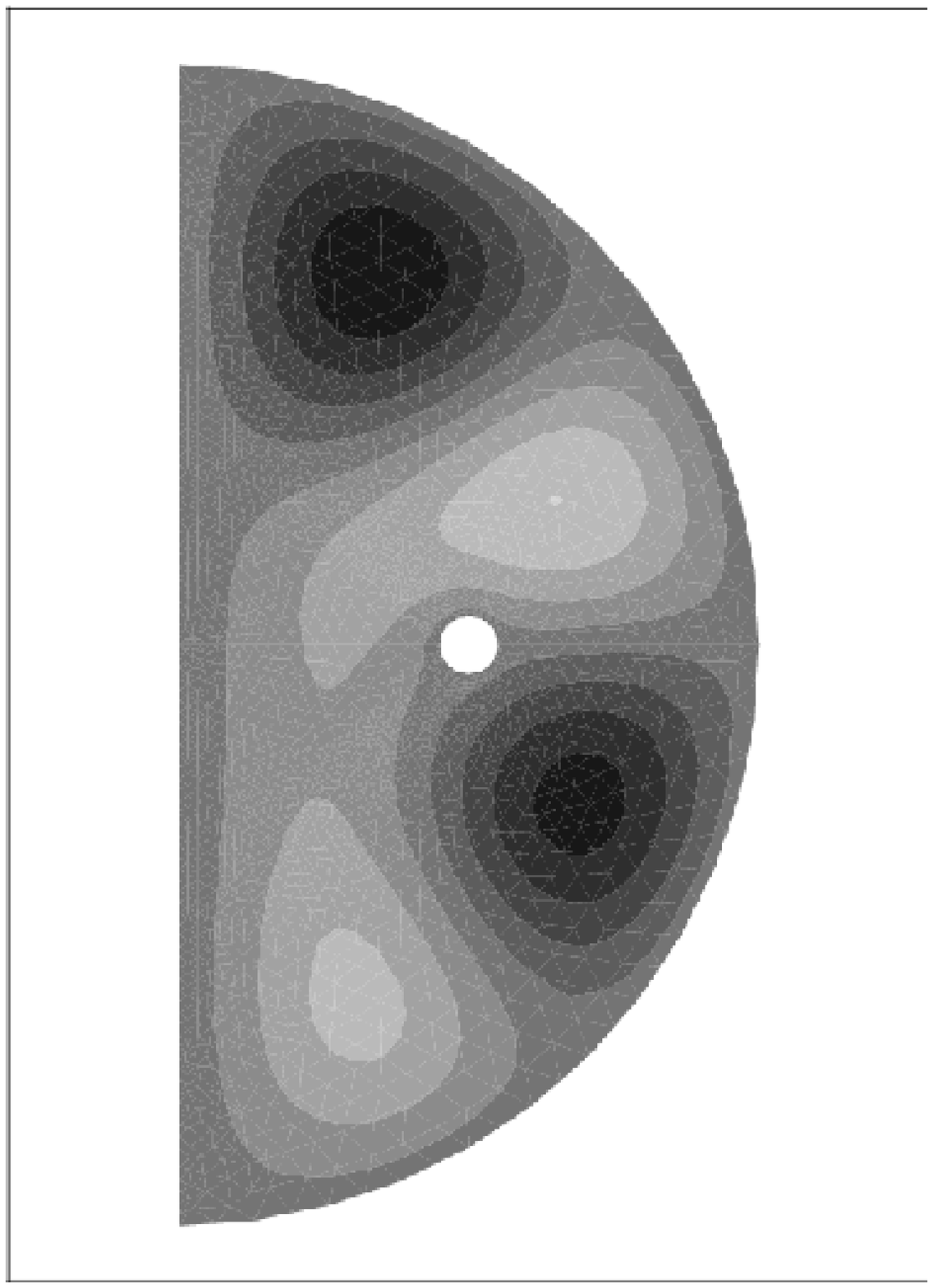}
\hspace{.06cm}\includegraphics[width=3.1cm,height=4.4cm]{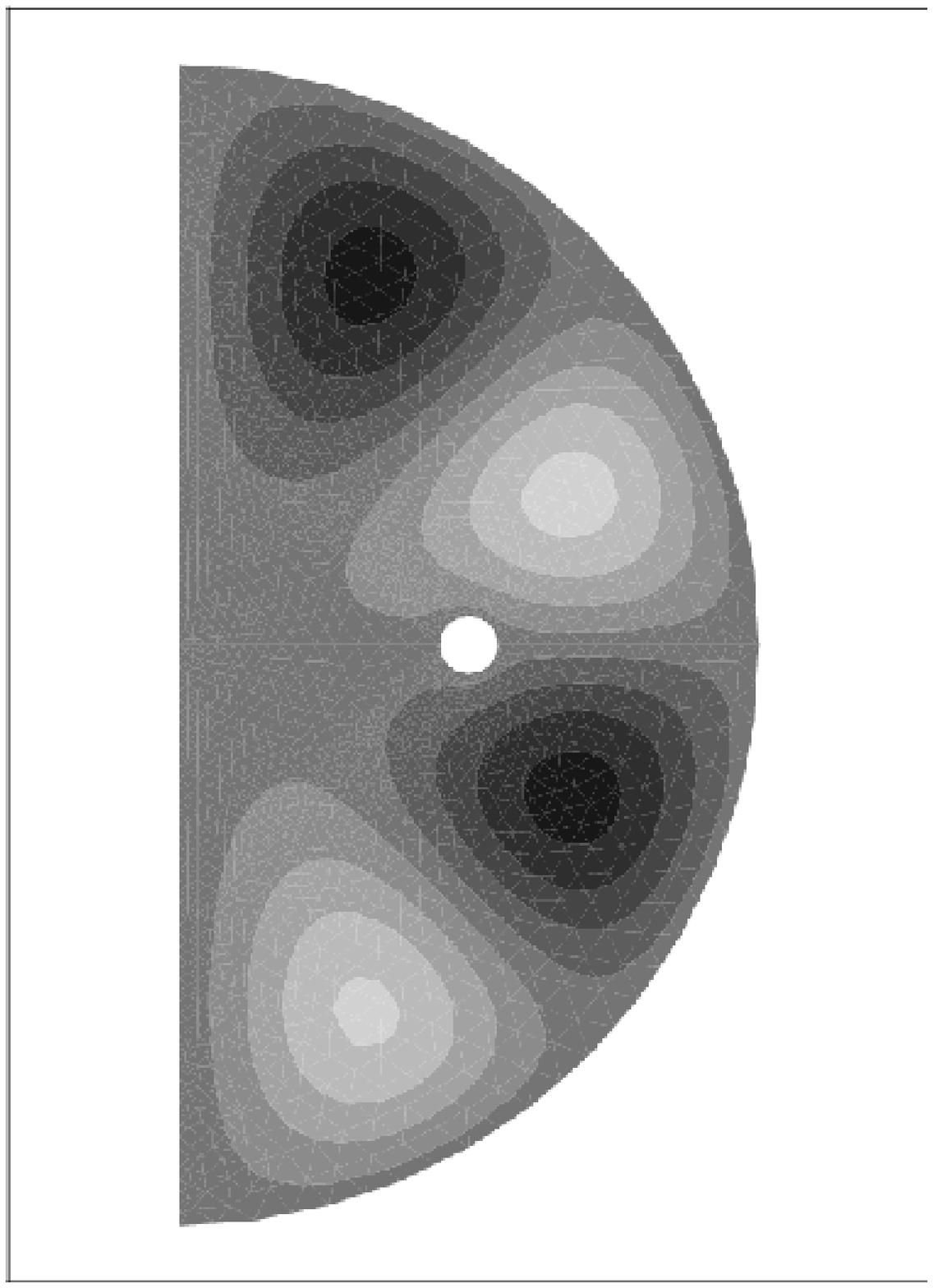}}\vspace{.2cm}
\vspace*{8pt}
\caption{\small Rotating wave in a Hill's spherical vortex: case of the domain $B$.
The sequence is referred to half cycle and terminates with the colors inverted. The small inner
circle is spinning at a frequency more than 12 times greater, dragging the spherical wave along
circulating paths.}
\end{figure}
\end{center}

\begin{center}
\begin{figure}[!ph]
\centerline{\includegraphics[width=3.1cm,height=4.4cm]{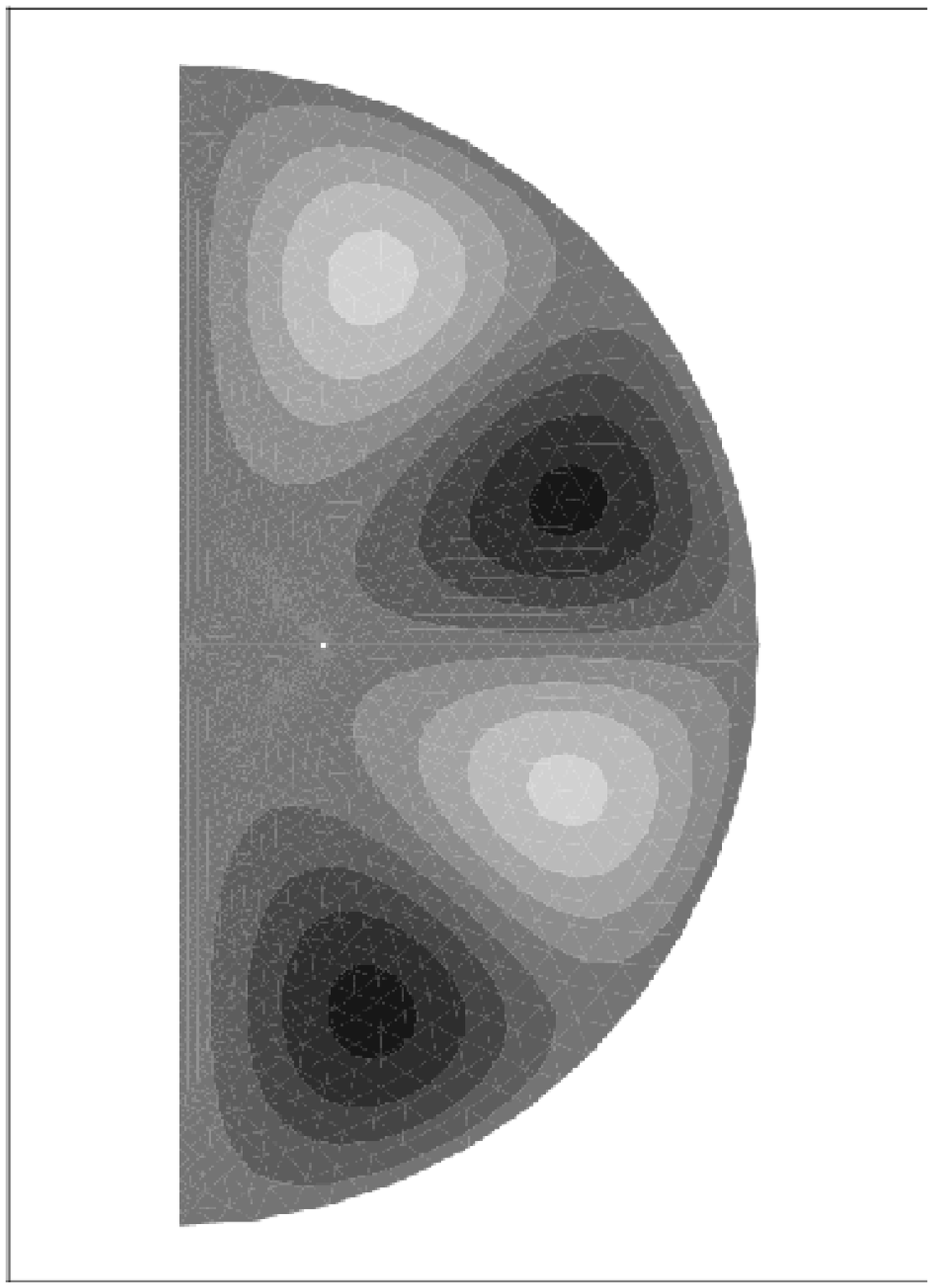}
\hspace{.06cm}\includegraphics[width=3.1cm,height=4.4cm]{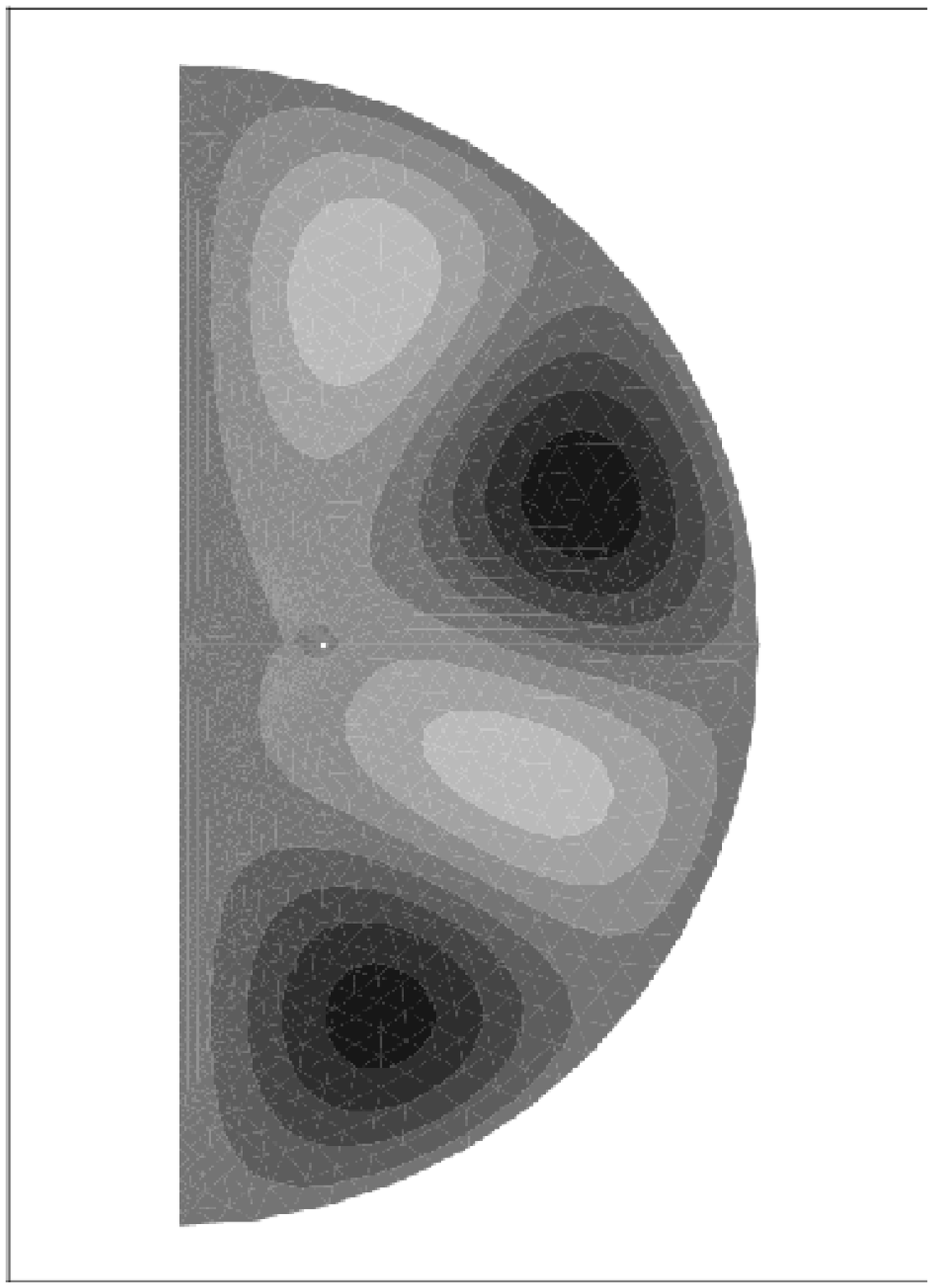}
\hspace{.06cm}\includegraphics[width=3.1cm,height=4.4cm]{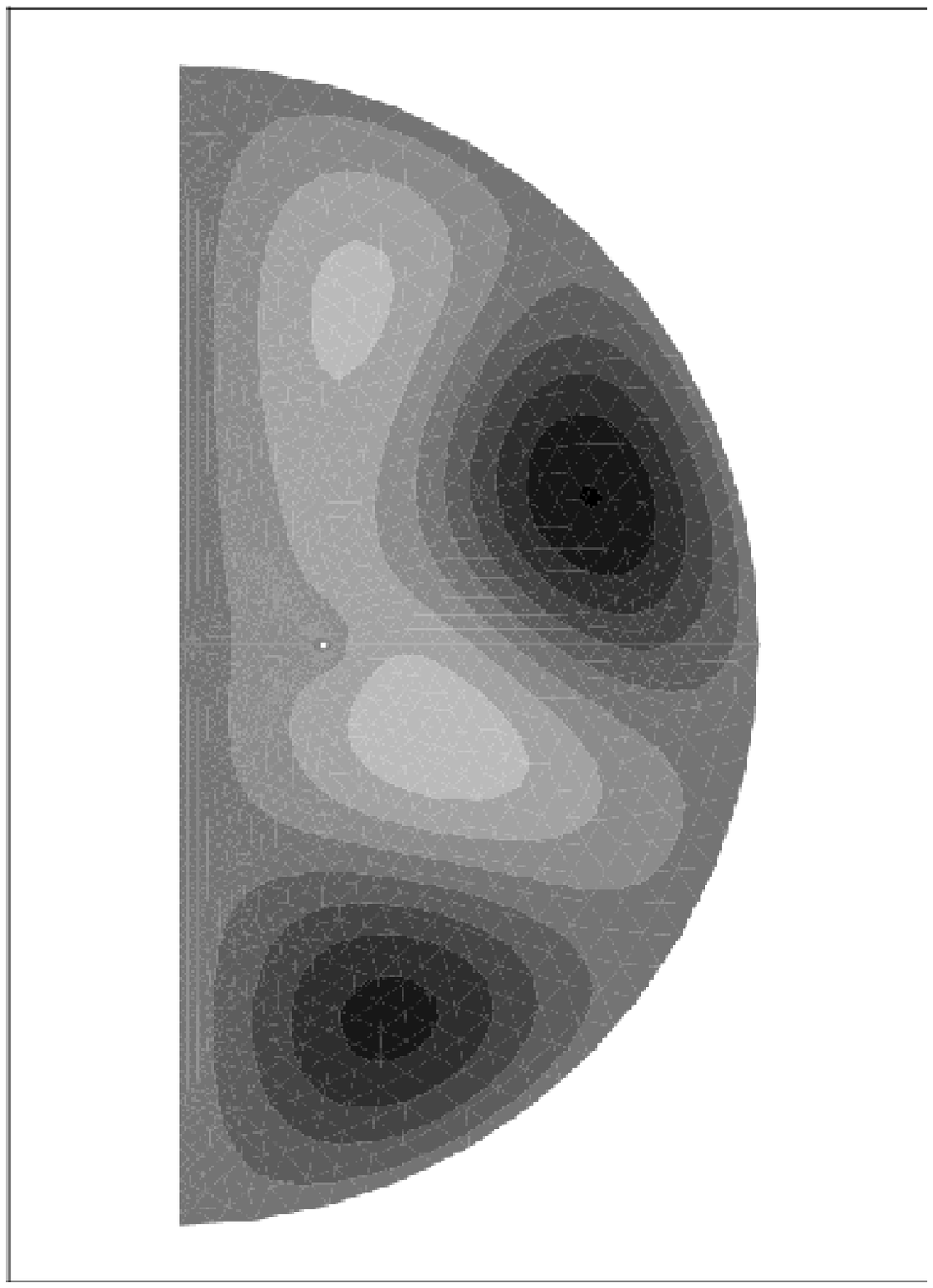}
\hspace{.06cm}\includegraphics[width=3.1cm,height=4.4cm]{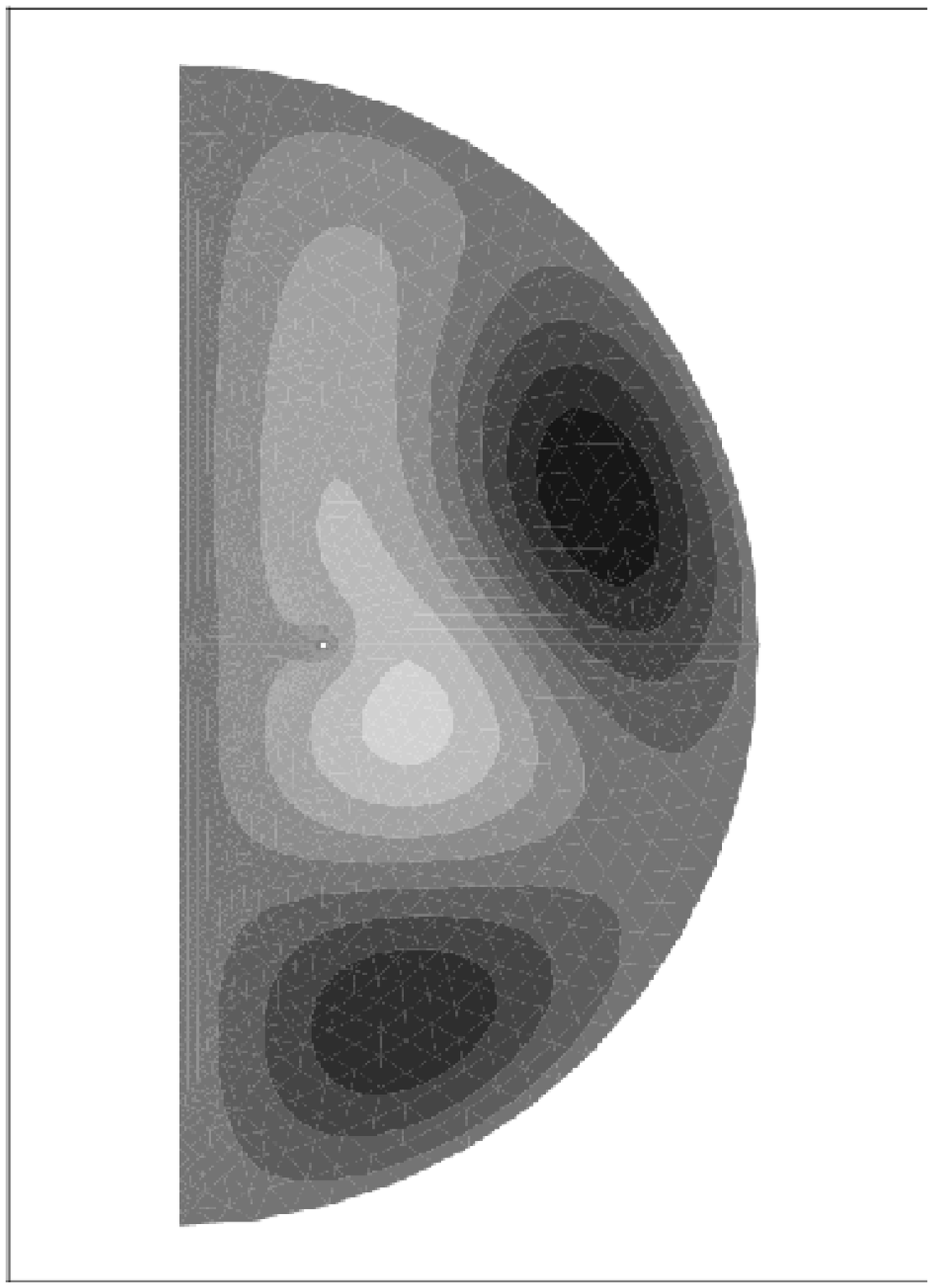}} \vspace{.07cm}
\centerline{\includegraphics[width=3.1cm,height=4.4cm]{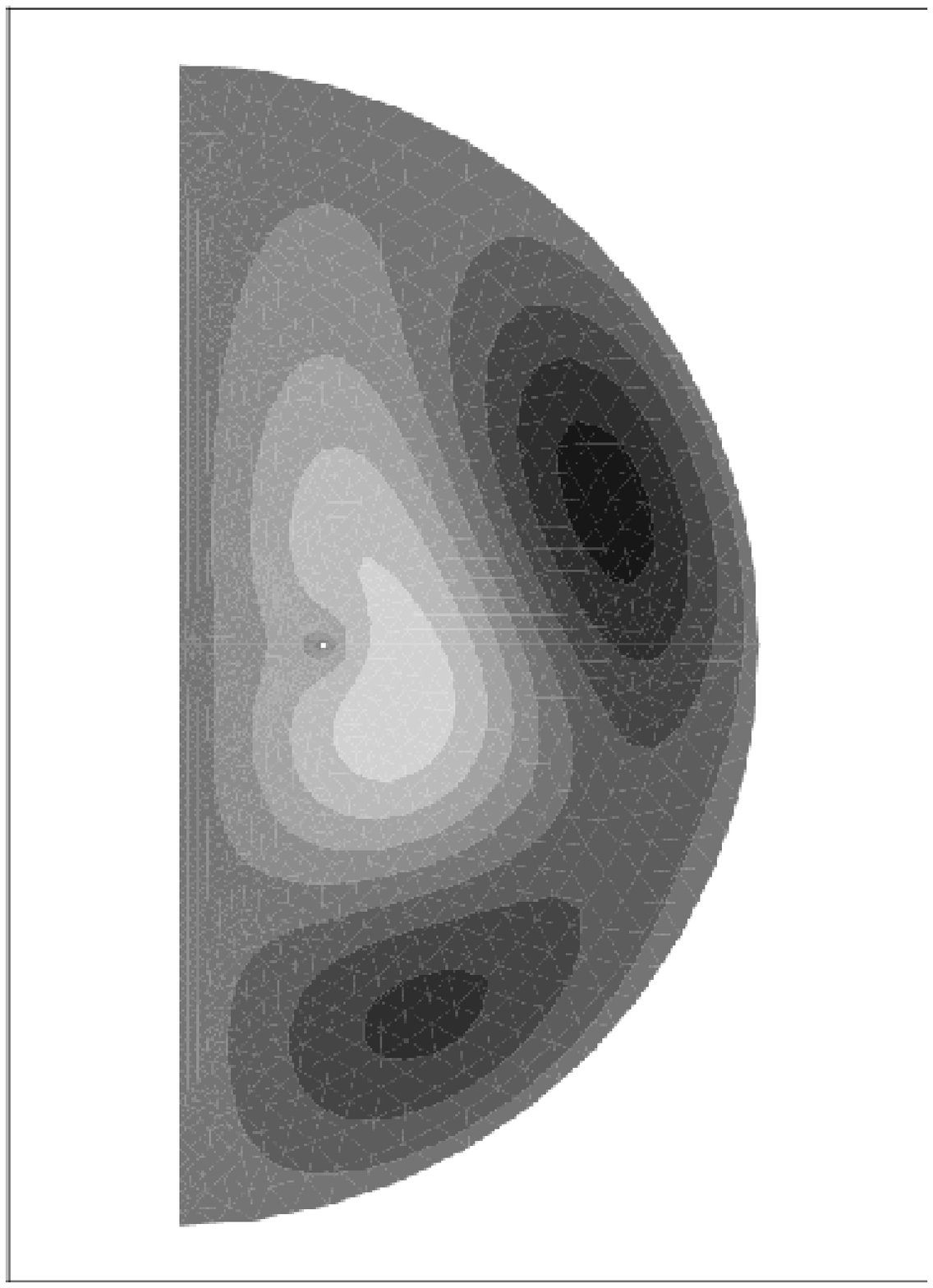}
\hspace{.06cm}\includegraphics[width=3.1cm,height=4.4cm]{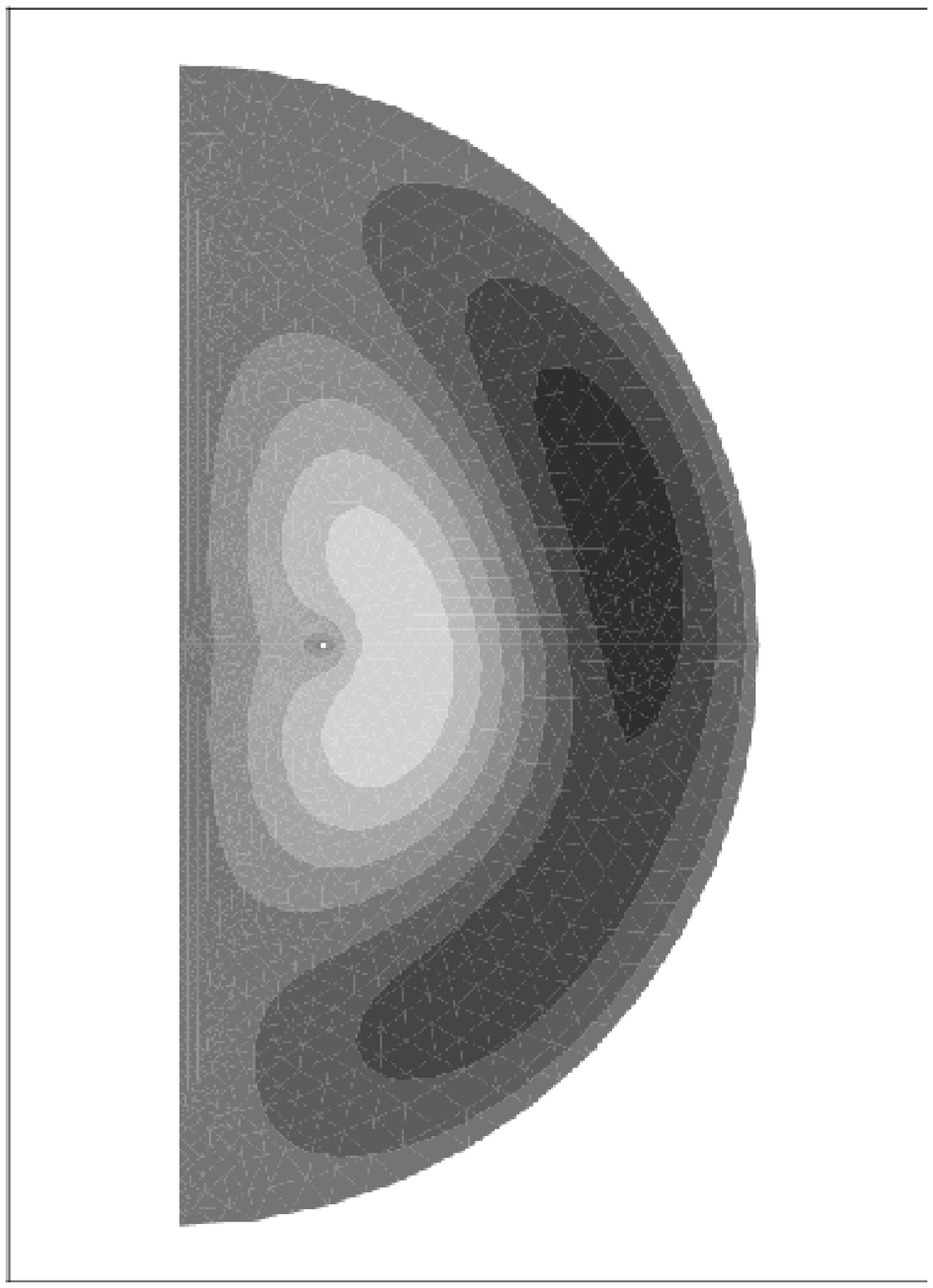}
\hspace{.06cm}\includegraphics[width=3.1cm,height=4.4cm]{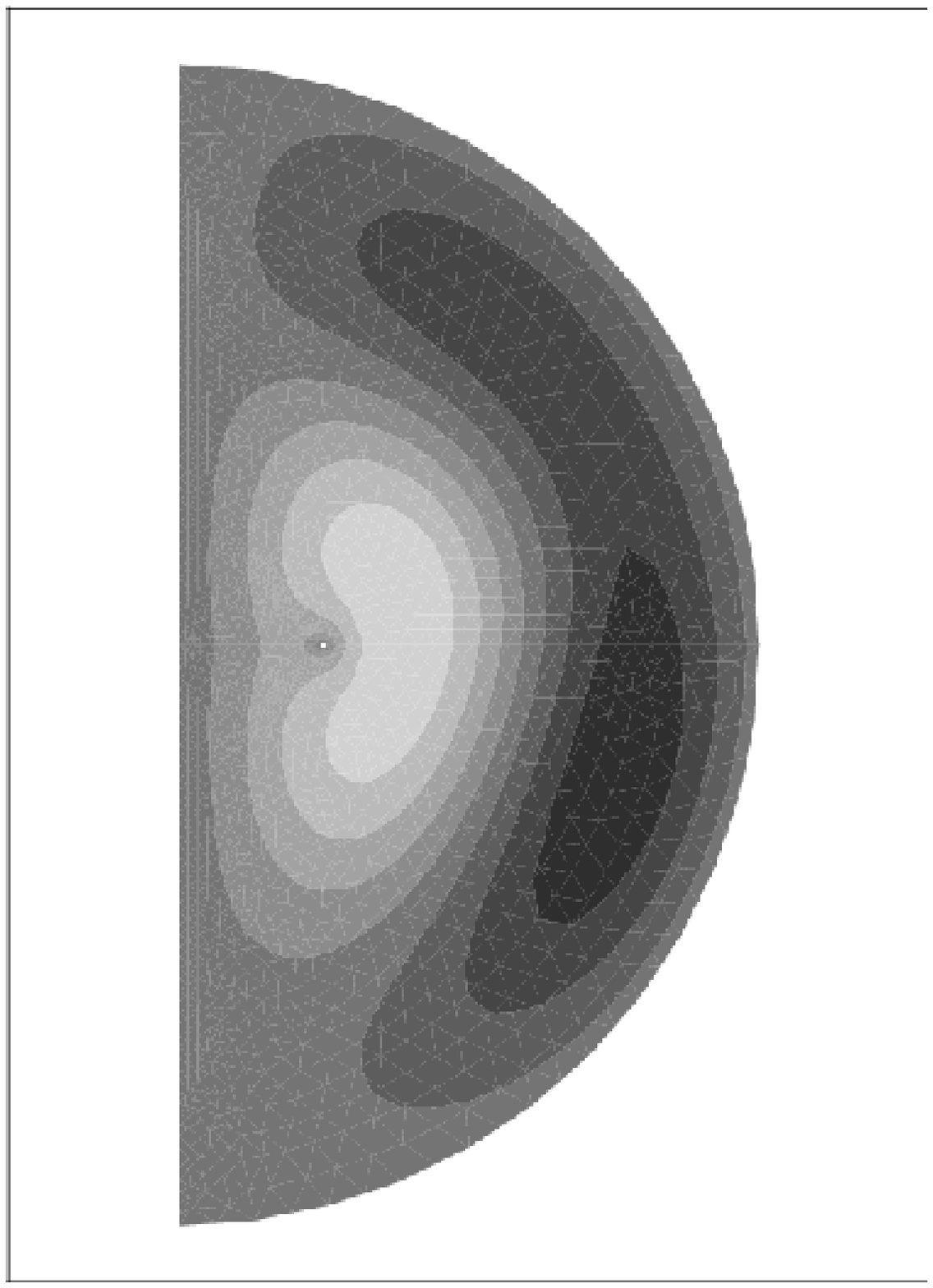}
\hspace{.06cm}\includegraphics[width=3.1cm,height=4.4cm]{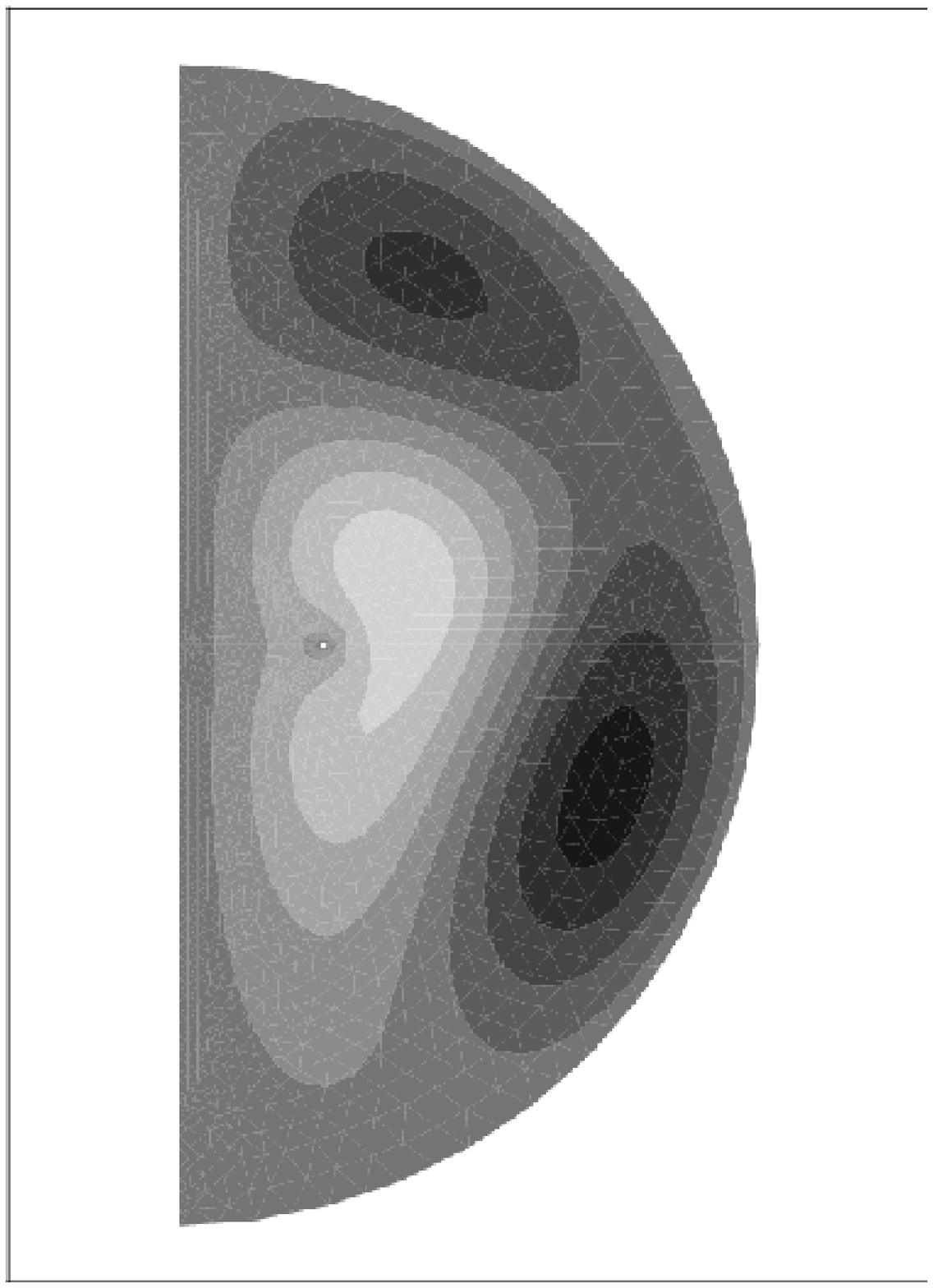}} \vspace{.07cm}
\centerline{\includegraphics[width=3.1cm,height=4.4cm]{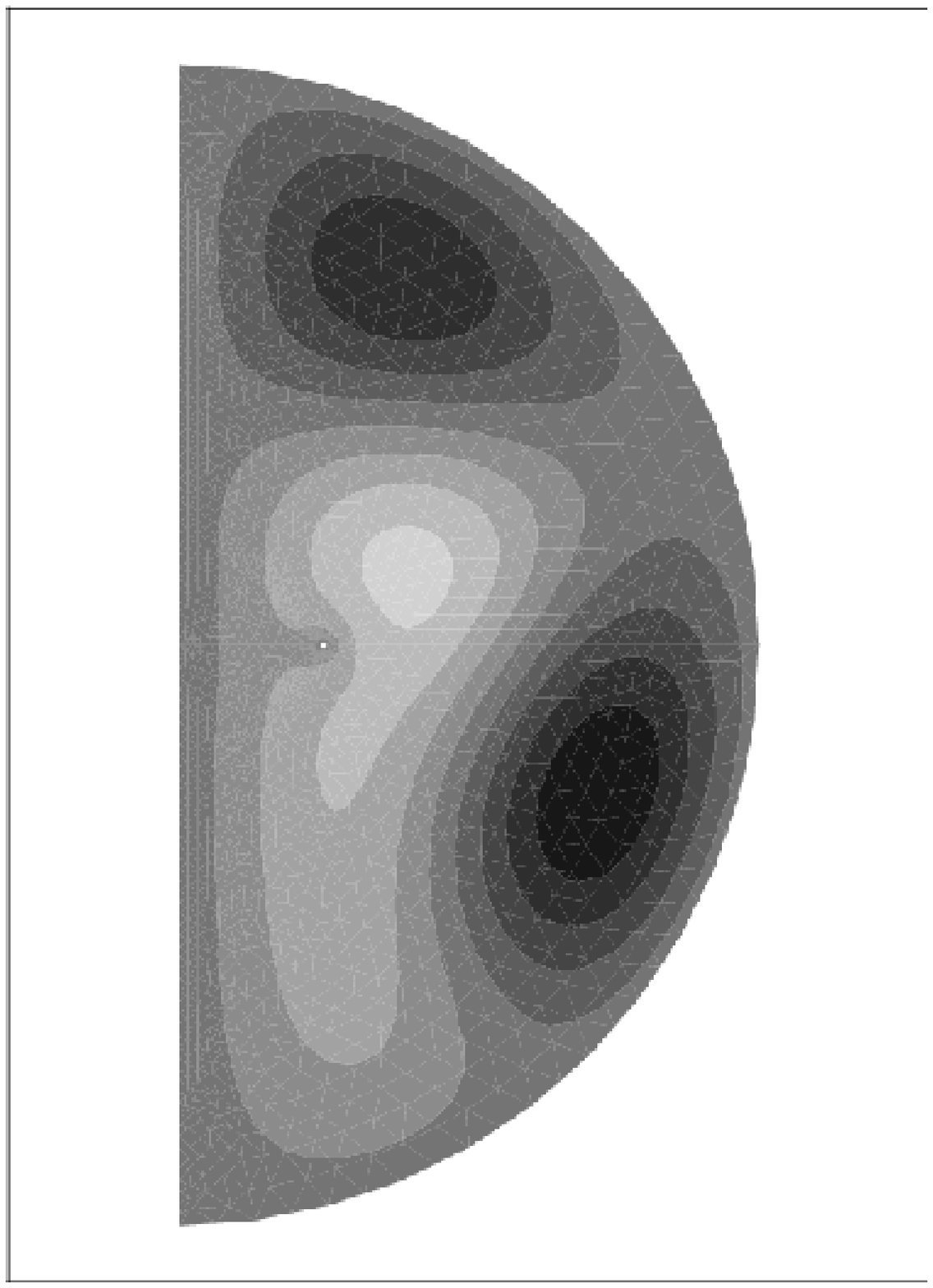}
\hspace{.06cm}\includegraphics[width=3.1cm,height=4.4cm]{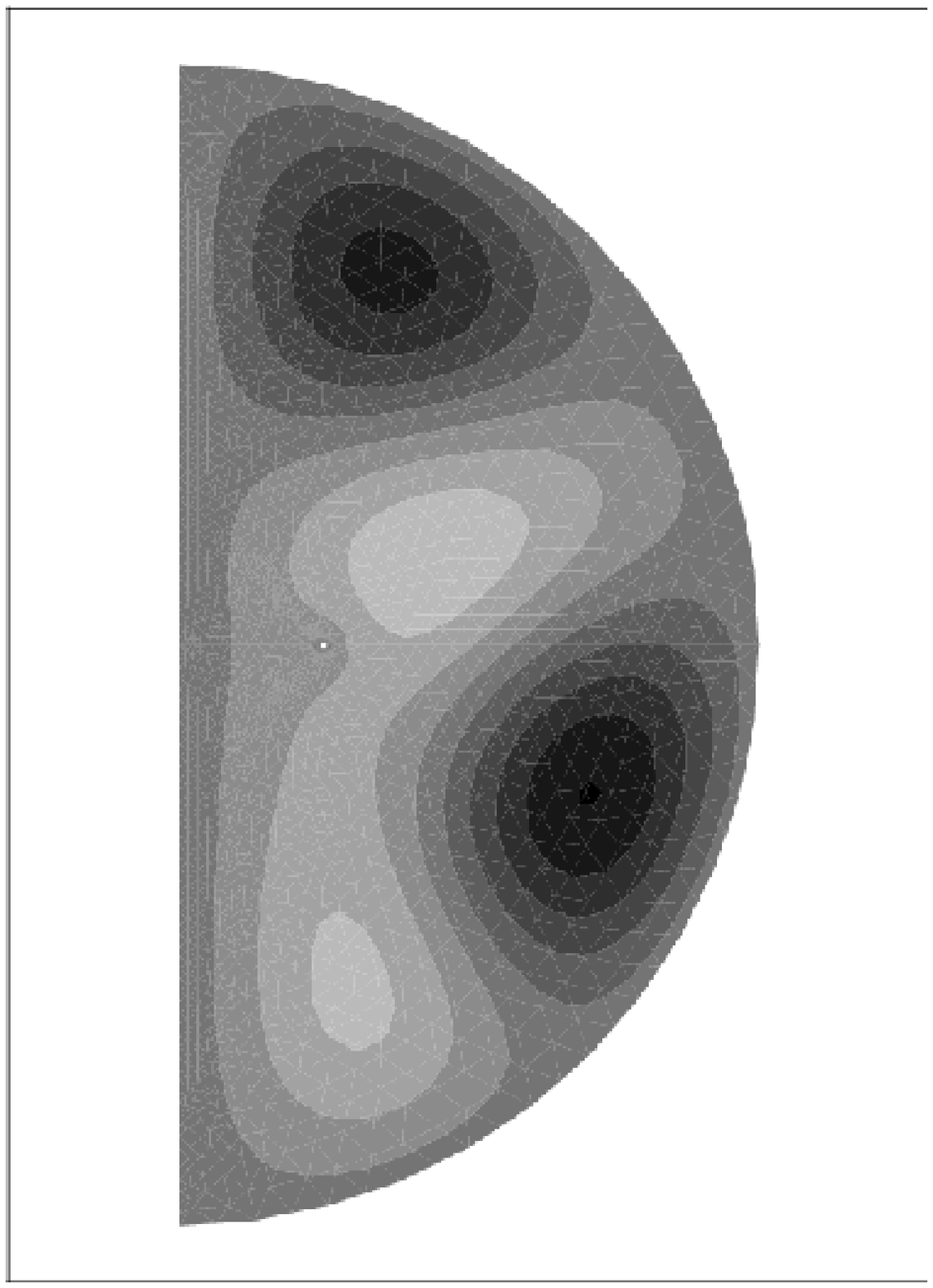}
\hspace{.06cm}\includegraphics[width=3.1cm,height=4.4cm]{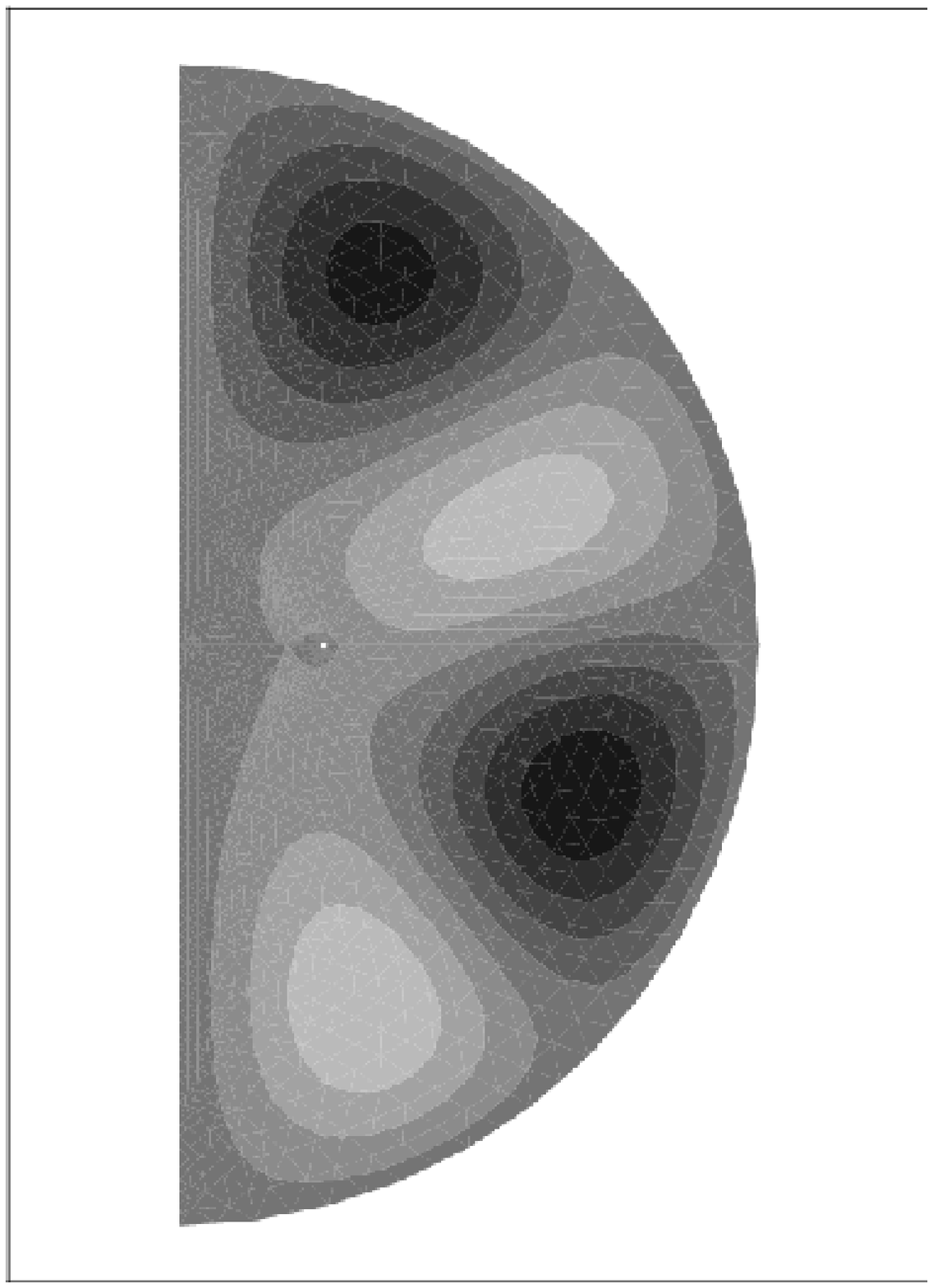}
\hspace{.06cm}\includegraphics[width=3.1cm,height=4.4cm]{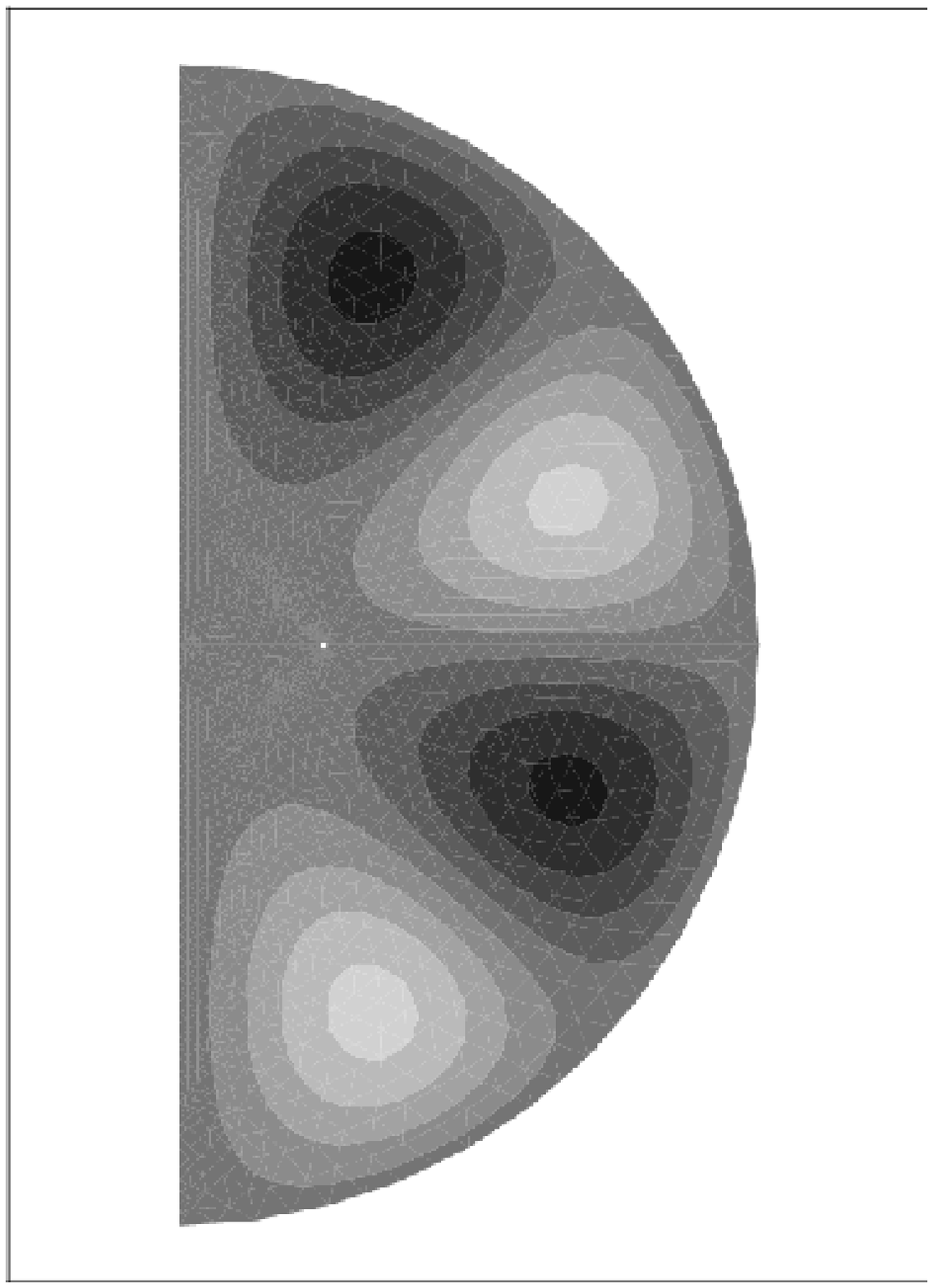}}\vspace{.2cm}
\vspace*{8pt}
\caption{\small Rotating wave in a Hill's spherical vortex: case of the domain $C$. The tiny
inner circle (hardly visible in the pictures) is spinning at a frequency about 120 times greater. The
behavior is qualitatively the same as in figure 8, but now the wave do not pass between the circle
and the vertical axis.}
\end{figure}
\end{center}

\newpage

From the size of the hole, we can estimate the frequency of
rotation of the internal toroid regions. Such a frequency is
proportional to the square root of a suitable eigenvalue. We take
the value $\sqrt{26.37}\approx 5.135$, that, according to table 1,
is the square root of the eigenvalue corresponding to the fourth
and the fifth eigenfunctions, for a  circle  of radius 1.
Afterwards, by scaling, we get the frequencies on the reduced
circles: $5.135/ r_B\approx 101.68$ and $5.135/ r_C\approx 987.5$.
Thus, the ratio between the frequencies of the inner spinning
rings and the spherical vortices are given by (see table 4, the
last line of the second and the third columns):
$101.68/\sqrt{68.15}\approx 12.31$ and $987.5/\sqrt{67.17}\approx
120.48$, respectively. Note that these numbers may be affected by
rounding errors (in particular, we expect an error bound of $\pm
0.005$ in the computation of the eigenvalues), therefore they
should be taken with a little caution. Of course, more trustable
results can be obtained with a finer mesh.
\par\smallskip

Finally, considering the second picture of figure 5,  we can study
the next external spherical shell, whose section is given by the
set $D$ of figure 6. Let us suppose that the radius of the inner
circumference is equal to 1, and let us find the outer radius
$r_D$ in order to have $\lambda_4 = \lambda_5$. From the
experiments we deduce  $r_D\approx 5.80$. This gives $\lambda_4 =
\lambda_5\approx 1.99$ (see the last column of table 4).  The
number of degrees of freedom, including internal and boundary
points, is 1857. The two corresponding eigenfunctions are shown in
figure 10. Their time evolution is very similar to the one of
figure 9.

\par\smallskip

We can confirm and improve these last results by  explicitly
determining, in terms of classical orthogonal basis, the solutions
of the wave equation. Due to the simplicity of the domain $D$, we can
actually use separation of variables in spherical coordinates
$(r,\theta ,\phi)$. Note that here the variable $r$ has a
different meaning: before it was the distance from the axis
$(0,z,0)$, now it is the distance from the center $(0,0,0)$.
According to Ref.~\cite{watson} and Ref.~\cite{funaro}, p. 15, we can get
solutions to the spherical vector wave equation by linear
combination of the functions:
$$
\sqrt{r}~J_{n+1/2}(r \sqrt{\lambda})~\sin\theta~P^\prime_n(\cos\theta )~\cos(ct\sqrt{\lambda})
$$
\begin{equation}\label{eq:besselb}
\sqrt{r}~Y_{n+1/2}(r \sqrt{\lambda})~\sin\theta~P^\prime_n(\cos\theta )~\cos(ct\sqrt{\lambda})
\end{equation}
where $P_n$ is the $n$-degree Legendre polynomial and $J_{n+1/2}$ and $Y_{n+1/2}$ are the Bessel
functions of the first and the second kind, respectively.

\begin{center}
\begin{figure}[h]
\centerline{\includegraphics[width=6.cm,height=8.5cm]{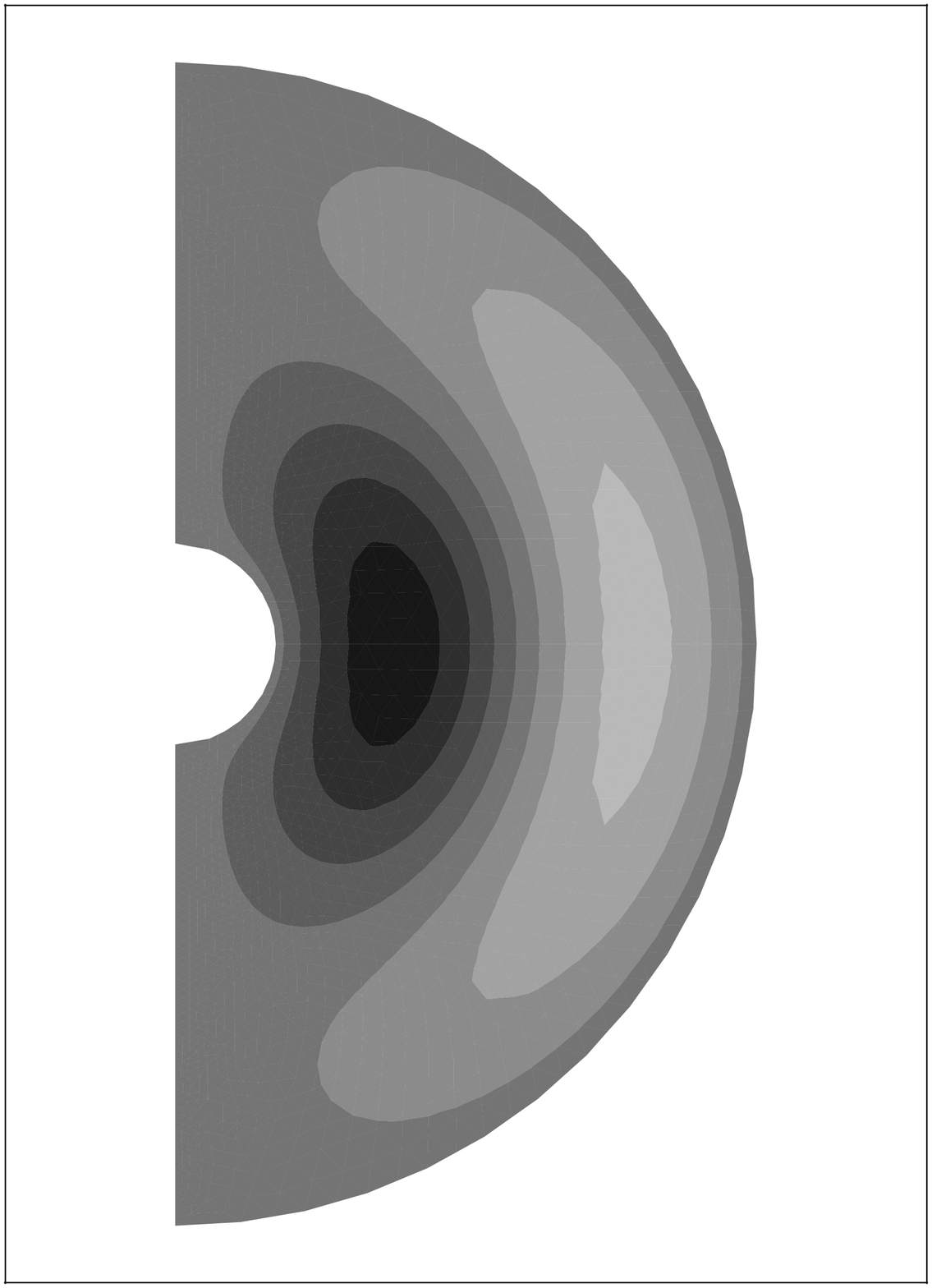}
\hspace{.2cm}\includegraphics[width=6.cm,height=8.5cm]{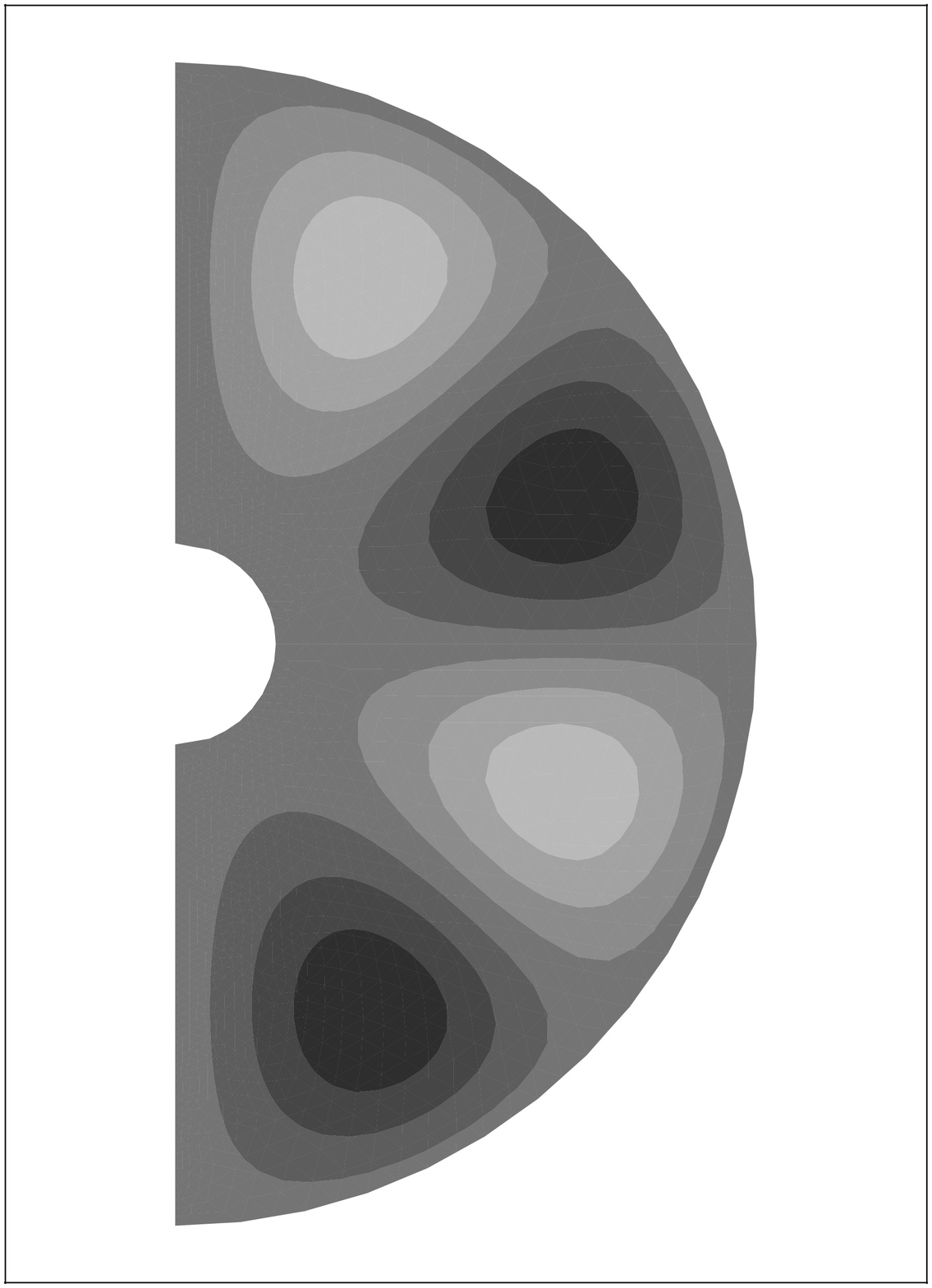} }
 \vspace*{8pt}
\caption{\small Two independent eigenfunctions in the case of the domain
$D$, for $\lambda_4 = \lambda_5\approx 1.9639$.}
\end{figure}
\end{center}

We are concerned with the cases $n=1$ and $n=4$. For these values,
the functions in (\ref{eq:besselb}) have, with respect to the
azimuthal variable $\theta$, one single bump or 4 consecutive
bumps, respectively (see figure 10). As a matter of fact, we have:
$\sin\theta P^\prime_1(\cos\theta )=\sin\theta~$ and $~\sin\theta
P^\prime_4(\cos\theta )=\frac{5}{2}\sin\theta\cos\theta
(7\cos^2\theta-3)$. Regarding instead the radial variable $r$, we
would like to find  suitable linear combinations of the functions
in (\ref{eq:besselb}), in order to impose homogeneous Dirichlet
boundary conditions at $r=1$ and $r=r_D$.  In agreement with the
results obtained by implementing the finite element method,
playing with the zeros of Bessel functions, yields:
\begin{equation}\label{eq:ribes}
\lambda =\lambda_4 =\lambda_5~\approx~1.9639~~~~~~~~{\rm and}~~~~~~~~~~r_D~\approx~5.839
\end{equation}
These data are now more accurate. The correct assemblage of the
radial basis  functions can be seen in figure 11. The first picture
shows a linear combination of $J_{3/2}$ and $Y_{3/2}$ ($n=1$), the
second one a linear combination of  $J_{9/2}$ and  $Y_{9/2}$
($n=4$). Up to multiplicative constants, this setting is unique if
we impose the boundary conditions at the endpoints of the same
interval $[1,r_D]$. In a similar way, we can also update the
values of table 4 (fourth column), namely: $\lambda_1\approx
.626$, $\lambda_2\approx .980$, $\lambda_3\approx 1.432$.

\begin{center}
\begin{figure}[h]
\vspace{-2.cm}
\hspace{-.4cm}
\centerline{
\includegraphics[width=4.2cm,height=6cm]{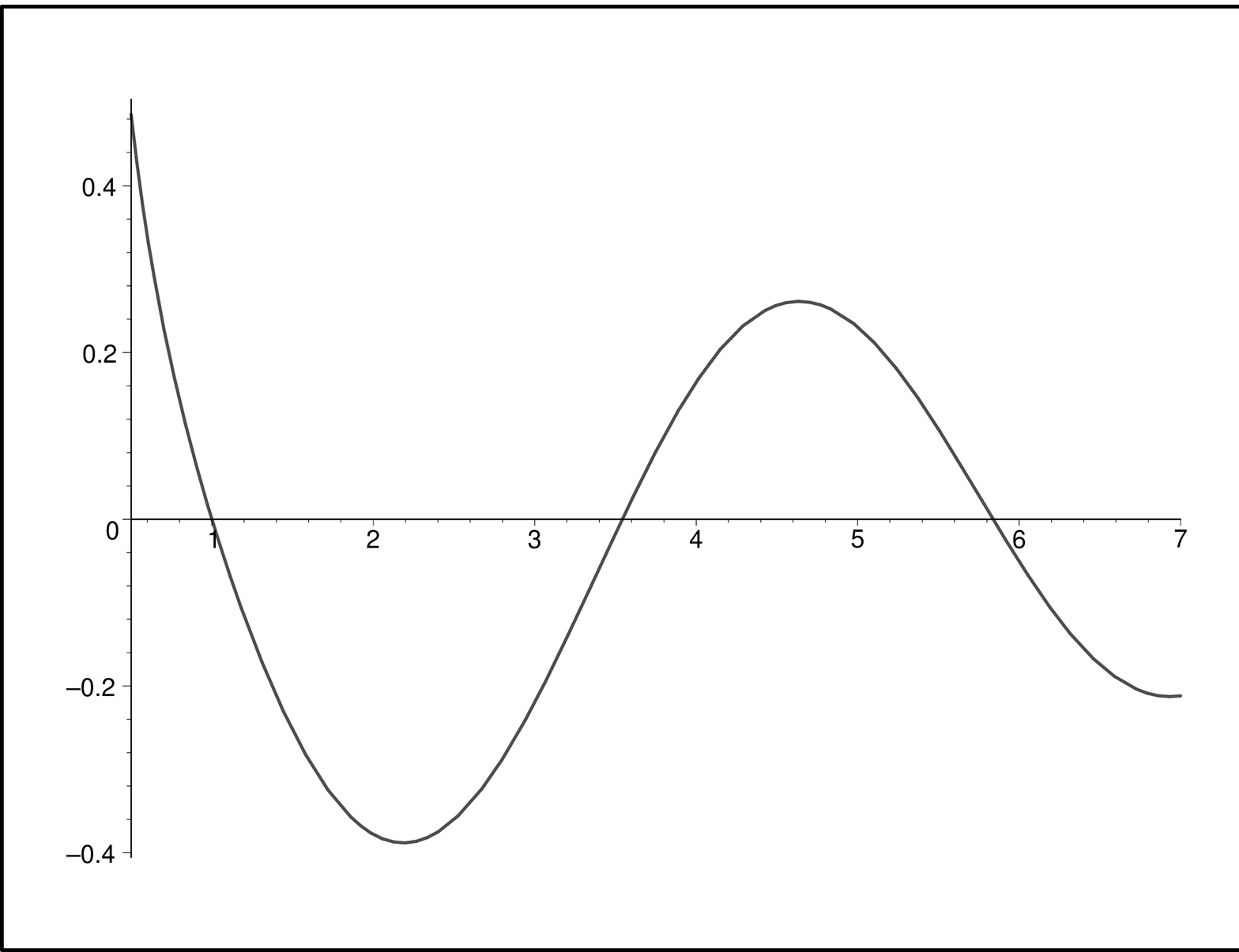} 
\hspace{2.truecm}
\includegraphics[width=4.2cm,height=6cm]{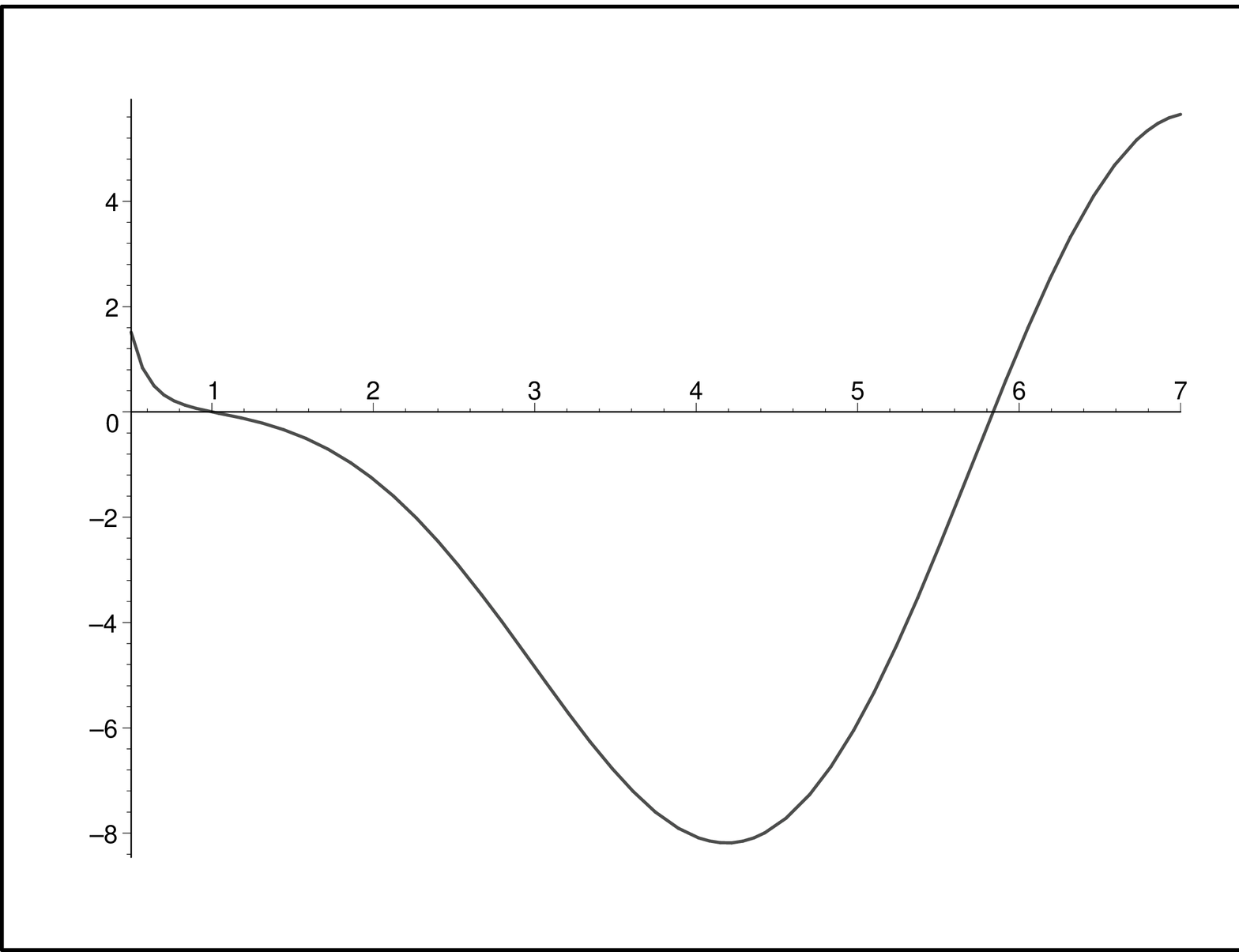}
}
\vspace{1.cm}
\caption{\small Plots of combinations of Bessel functions of  first and
second kind, vanishing at $r=1$ and $r=r_D$.}
\end{figure}
\end{center}

Further external shells are obtained by linearly amplifying the
domain $D$. In this way, the size of the successive nested shells
grows geometrically by a factor $5.839$, and the corresponding
frequencies are progressively reduced by the same  factor. Like
gears of increasing magnitude connected together, the shells
transmit their signals far away from the source, through a
quantized process, causing a decay of the frequency at each step
(see Ref.~\cite{funaro}, section 6.1).
\par\smallskip

As a last example, we examine the case of figure 12, where the
curved part of the domain $E$ is an ellipse. We recall that when
the Hill's vortex is perfectly spherical (domain $A$ in figure 6)
there is no chance to get coincident eigenvalues (see the first
column of table 4). However, by reducing the vertical axis, the
situation improves without the help of internal holes. We found
out that $\lambda_4=\lambda_5$ when the ratio between the vertical
and the horizontal axis is equal to .6035 (see table 5). The
evolution of the solution is expected to be similar to that of
figures 8 and 9.
\par\smallskip

In Ref.~\cite{sullivan}, predictions and experimental verifications
about the formation of stable configurations consisting of an
elliptic-shaped Hill's type vortex, generated by an internal thin
vortex ring, are discussed. Further indications might come from
our approach, based on the study of $\lambda_4$ and $\lambda_5$,
by simultaneously arranging the eccentricity of the ellipse and
the location and the size of the internal vortex. The
configuration is going to be similar to the one of figure 4 in
Ref.~\cite{sullivan}. We tried a few qualitative experiments in this
direction, but, having too many degrees of freedom, a careful
analysis is a technical exercise that we would prefer to avoid at
the moment. Therefore, we do not anticipate any result.

\begin{figure}[t]
\centerline{\hspace{1.cm}{\includegraphics[width=6cm,height=6cm]{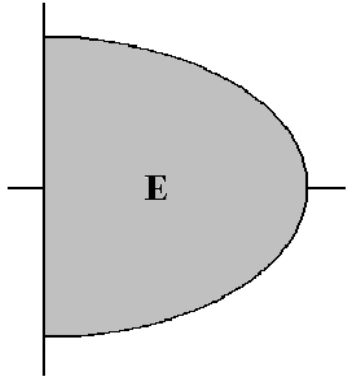}}}
\vspace*{8pt}
\caption{\small Elliptic section with semi-axis equal to 1 and .6035.}
\end{figure}

\begin{table}\vspace{.5cm}
\label{tab1} \noindent \[
\begin{array}{|c|c|}
  \hline
   &   ~~{domain}~E  ~~  \\
  \hline
  \lambda_1 & ~~26.96  \\
  \lambda_2 & ~~56.97  \\
  \lambda_3 & ~~64.53  \\
  \lambda_4 & 101.95  \\
  \lambda_5 & 101.99  \\
    \hline
\end{array}
\]
\begin{caption}{\small Eigenvalues corresponding to the domain E of figure 12.}
\end{caption}
\end{table}

\medskip

\section{Conclusions}

With the help of numerical simulations, we demonstrated the
possibility of building electromagnetic waves trapped in bounded
3-D regions of space. We did not insist too much on issues related to
the performances of the algorithms. We are conscious of the fact 
that the methods used can be certainly ameliorated, in terms
of costs versus accuracy. Nevertheless, this was not our primary concern.
\par\smallskip

We think that the results we got may have a
general validity independently on the field of applications.
In fact, the approach we followed, based on the determination
of periodic solutions to the wave equation through the analysis of
certain eigenvalues of a suitable differential operator, may
be applied in several circumstances. For example, as an alternative,
this technique may be employed in fluid dynamics (for stable periodic flows), 
where computations are usually
carried out by discretizing the equations by some time-advancing
procedure. However, the meaning of the results obtained here is deeper,
since they are related to the approximation of a complete system of hyperbolic
equations (namely the Maxwell's equations) and not just to the detection of
the flow-field. As a matter of fact, the information carried by
our waves is not a scalar density field, but it consists of two separated 
vector fields: the electric and the magnetic ones. These can be fully 
expressed by the relations in (\ref{eq:camp}).
\par\smallskip

A recent subject of research, that could benefit from our
investigation, is the detection of quantum vortex rings in
superfluid helium (the literature in the field is very rich, see
for instance Ref.~\cite{barenghi} for a general overview). Note that,
in liquid helium, the thickness of these rings is on the order
of a few Angstroms. Other related topics might be the study of
{\sl ball lightning} phenomena (see for instance Ref.~\cite{alana} and
Ref.~\cite{zou}) or {\sl ring nebulae} and their halos (see for
instance Ref.~\cite{bryce}). 

\par\smallskip 

Our analysis might inspire interesting applications, that
at the moment we are unable to predict. Indeed, we showed that it is
possible to detect 3-D regions of space, where an electromagnetic 
wave, suitably supplied through the boundary, can freely travel
without bouncing off the walls. Moreover, we also explained how to build
these regions. We are not enough experienced in the field of applications
to find an employment for such resonant boxes, but this paper may suggest to some 
interested reader how to use them to create new tools or improve existing ones.
\par\smallskip  

Finally, we recall that the stability of these
structures is a consequence of gravitational modifications of the
space-time generated by the movement of the wave itself, according
to Einstein's equation. In fact, the wave travels like a fluid,
along closed stream-lines, that turn out to be geodesics of such a
modified geometry (see also Ref.~\cite{funaro2}).  Thus, in the electrodynamic 
framework, a complete analysis of stability depends on the resolution of a
complex relativistic problem. Nevertheless, we hope that the
simple investigation carried out in this paper may represent a
step ahead towards a better comprehension of the mechanism of
vortex formation, whatever the constituents are (fields or real
matter).
\par
\end{document}